\documentclass[pre,amsmath,amssymb,reprint]{revtex4-1}

\usepackage[utf8]{inputenc}

\usepackage{color}
\usepackage[draft]{changes}
\usepackage{amsfonts,MnSymbol,bm}
\usepackage{graphicx}

\newcommand{\1}{\Bar{1}}
\newcommand{\2}{\Bar{\Bar{1}}}

\newcommand{\mb}[1]{\mbox{\bfseries \itshape #1}}

\begin{document}

\title{A taxonomy for generalized synchronization between flat-coupled systems}

\author{Christophe Letellier}
\email{Christophe.Letellier@univ-rouen.fr}

\affiliation{Rouen Normandie University --- CORIA, Avenue de l'Universit\'e,
F-76800 Saint-Etienne du Rouvray, France}

\author{Ludovico Minati}
\email{lminati@uestc.edu.cn, ludovico.minati@unitn.it}

\affiliation{School of Life Science and Technology, University of Electronic
Science and Technology of China, Chengdu, China}
\affiliation{Center for Mind/Brain Science (CIMeC), University of Trento, 
38123 Trento, Italy}

\author{Irene Sendi\~{n}a Nadal}
\email{irene.sendina@urjc.es}

\author{I. Leyva}
\email{inmaculada.leyva@gmail.com}

\affiliation{Complex Systems Group \& GISC, Universidad Rey Juan Carlos,
28933 M\'ostoles, Madrid, Spain}

\affiliation{Center for Biomedical Technology, Universidad Polit\'ecnica de
Madrid, 28223 Pozuelo de Alarc\'on, Madrid, Spain}

\date{\today, Submitted to {\it Physical Review E}}

\begin{abstract}
Generalized synchronization is plausibly the most complex form of 
synchronization. Previous studies have revealed the existence of weak or strong 
forms of generalized synchronization depending on the multi- or mono-valued 
nature of the mapping between the attractors of two unidirectionally-coupled 
systems. Generalized synchronization is here obtained by coupling two systems 
with a flat control law. Here, we demonstrate that the corresponding 
first-return maps can be topologically conjugate in some cases. Conversely, the 
response map can foliated while the drive map is not. We describe the 
corresponding types of generalized synchronization, explicitly focusing on the 
influence of the coupling strength when significantly different dimensions or 
dissipation properties characterize the coupled systems. A taxonomy of 
generalized synchronization based on these properties is proposed.
\end{abstract}

\maketitle

\section{Introduction}

Synchronization was already a phenomenon commonly investigated during the early
development of nonlinear electronic circuits in the 1920s 
\cite{Vin19,App22,Hun47}. Once chaotic systems became popular 
\cite{Lor63,Ros76c,Mat85,Hin84}, it was shown that two copies of such systems 
could also be synchronized despite their extraordinary sensitivity to initial 
conditions \cite{Pec90,Pik03,Hua09}. Soon afterward, different types of 
synchronization were distinguished: complete synchronization \cite{Pec90},
phase synchronization \cite{Ros96}, lag synchronization \cite{Ros97}, among 
others \cite{Boc02}. One way to detect synchronization is to plot the two 
coupled elementary variables against each other. Lissajous introduced this 
technique for investigating vibrating motion \cite{Lis57}, and it was later 
applied to electrical variables by Appleton \cite{App22} and Huntoon 
\cite{Hun47}. The Lissajous curve is a straight 
line for complete synchronization, whereas it is an ellipse when the two systems produce lag-synchronized period-1 orbits. These types of synchronization have been commonly 
observed between structurally identical oscillators, that is, having the same 
governing equations but a parameter mismatch. Whenever
the two coupled systems are structurally different, getting one of these three 
synchronization types is impossible since, even if they can share a 
given variable, the others are necessarily different due to the different 
couplings among the variables. A form of generalized synchronization  is 
obtained instead, with most examples in the literature corresponding to  
unidirectional coupling (master-slave) schemes. In these cases, the Lissajous 
curve often describes a complex invariant set \cite{Rul95} resembling, for 
instance, a plane projection of a chaotic attractor. 

Let us consider two unidirectionally coupled systems. There is, thus, a drive 
system $\dot{\mb{x}}_{\rm d} = \mb{f}_{\rm d} (\mb{x}_{\rm d})$ whose 
trajectory ${\cal T}_{\rm d} \equiv \{ \mb{x}_{\rm d} (t) \}_{t > t_0} \in 
{\cal A}_{\rm d} \subset \mathbb{R}^{d_{\rm d}}$ is within an invariant set
${\cal A}_{\rm d}$. There is also a response system
$\dot{\mb{x}}_{\rm r} = \mb{f}_{\rm r} (\mb{x}_{\rm r})$ whose trajectory
${\cal T}_{\rm r} \equiv \{ \mb{x}_{\rm r} (t) \}_{t > t_0} \in 
{\cal A}_{\rm r} \subset \mathbb{R}^{d_{\rm r}}$ is within an invariant set
too when the coupling strength is null. Here we consider that $\mb{f}_{\rm d}
\neq \mb{f}_{\rm r}$, that is, the drive and response systems are structurally
different, therefore ${\cal A}_{\rm d} \neq {\cal A}_{\rm r}$ and 
$d_{\rm r}$ and $d_{\rm d}$ are not necessarily equal. In that case, when the 
coupling strength is null, there is no map $G$ such that ${\cal A}_{\rm r}
= G ({\cal A}_{\rm d})$. The trajectories ${\cal T}_{\rm d}$ and 
${\cal T}_{\rm r}$ are generalized synchronized if there exists a map $G$
such that ${\cal A}_{\rm r} = G ({\cal A}_{\rm d})$ and $G$ differs from
the identity, for all $(\mb{x}_{\rm d} (0), \mb{x}_{\rm r}(0)) \in 
{\cal B}_{\rm GS}$ where ${\cal B}_{\rm GS} \subset \mathbb{R}^{d_{\rm d}} 
+ \mathbb{R}^{d_{\rm r}}$ is the basin of generalized synchronization between
the coupled systems. When they are uncoupled, the systems here considered are 
such that the minimum embedding dimension ${\cal D}_{\rm s}$ of the 
corresponding dynamics is equal 
to the dimension of the system $d_{\rm s}$, that is, ${\cal D}_{\rm s} 
= d_{\rm s}$. This is no longer true when the coupling strength is not null.

Synchronization is often investigated when the two systems are coupled with 
a diffusive coupling term of the form $\kappa (s_{\rm r} - s_{\rm d})$ where 
$s_{\rm r} = h_{\rm r} (\mb{x}_{\rm r})$ and 
$s_{\rm d} = h_{\rm d} (\mb{x}_{\rm d})$ are the measured variables in the 
response and the drive systems, respectively, and $s_{\rm r}$ is the output of 
the response system. In such a case, the synchronization problem can be seen as 
a control problem. Since the actuating signal $u$ (the input) depends on the 
response system output, this is a closed-loop control (feedback, essentially 
equivalent to a proportional controller). When the 
drive and the response systems are coupled, the response system 
$\dot{\mb{x}}_{\rm r} = \mb{f}_{\rm r} (\mb{x}_{\rm r}, u)$ is therefore
controlled by the drive system. Since $\mb{f}_{\rm d}$ and $\mb{f}_{\rm r}$
are unidirectionally coupled, ${\cal T}_{\rm d} \in {\cal A}_{\rm d}$ as when 
there is no coupling. Conversely, the trajectory ${\cal T}_{\rm d}$ does not 
lie within ${\cal A}_{\rm r}$ but in another invariant set 
${\cal A}_{\rm r}^{\rm c}$. Thus, ${\cal A}_{\rm d} \in \mathbb{R}^{d_{\rm d}}$
and ${\cal A}_{\rm r}^{\rm c} \in \mathbb{R}^{d_{\rm d}} 
+ \mathbb{R}^{d_{\rm r}}$. When two variables $s_{\rm r}$ and $s_{\rm d}$ are
such as the Lissajous curve is a straight line, there is a simple relation 
between them and, typically, the dimension of the space in which 
${\cal A}_{\rm r}^{\rm c}$ takes place is reduced by, at least, one. 
If the coupling strength is sufficiently large, the invariant set 
${\cal A}_{\rm r}^{\rm c} \subset \mathbb{R}^{d_{\rm d}} (\mb{x}_{\rm r})$.
Note that $\mathbb{R}^{d_{\rm d}} (\mb{x}_{\rm r})$ is not spanned by the 
same axis than $\mathbb{R}^{d_{\rm d}} (\mb{x}_{\rm d})$. When $d_{\rm d} = 
d_{\rm r}$, there is no ambiguity regarding how the subspace 
$\mathbb{R}^{d_{\rm d}} (\mb{x}_{\rm r})$ can be interpreted: this is the 
space associated with the response system. When $d_{\rm r} > d_{\rm d}$, there 
are superfluous dimension and a subspace of $\mathbb{R}^{d_{\rm r}} 
(\mb{x}_{\rm r})$ is sufficient to embed the controlled response dynamics.
When $d_{\rm r} < d_{\rm d}$, some dimension must be added to the response
space $\mathbb{R}^{d_{\rm r}} (\mb{x}_{\rm r})$ for embedding the controlled
response dynamics. Consequently, the embedding dimension ${\cal D}_{\rm r}$ of 
the response dynamics is such that $d_{\rm d} \leqslant {\cal D}_{\rm r} 
\leqslant d_{\rm d} + d_{\rm r}$.

Since first termed by Rulkov and co-workers \cite{Rul95}, many examples of 
generalized synchronization have been reported in the literature 
\cite{Aba96,Joh98,Zha07,He09,Lym19}. One of the earliest examples was provided 
by Kocarev and Parlitz \cite{Koc96} between a drive R\"ossler system 
\cite{Ros76c}
\begin{equation}
  \left\{
    \begin{array}{l}
      \dot{x}_{\rm d} = - y_{\rm d} - z_{\rm d} \\[0.1cm]
      \dot{y}_{\rm d} =  x_{\rm d} + a y_{\rm d} \\[0.1cm]
      \dot{z}_{\rm d} = b + z_{\rm d} (x_{\rm d} - c)
    \end{array}
  \right.
\end{equation}
and a response Lorenz system \cite{Lor63}
\begin{equation}
  \left\{
    \begin{array}{l}
      \dot{x}_{\rm r} = - \sigma x_{\rm r} + \sigma y_{\rm r} \\[0.1cm]
      \dot{y}_{\rm r} =  R u - y_{\rm r} - u z_{\rm r} \\[0.1cm]
      \dot{z}_{\rm r} = - b z_{\rm r} + u y_{\rm r} 
    \end{array}
  \right.
\end{equation}
where the actuating signal $u = x_{\rm d} + y_{\rm d} + z_{\rm d}$. Here, the
response system is driven with an open-loop since $u$ does not depend on 
$\mb{x}_{\rm r}$. The response system is driven with its variable 
$x_{\rm r}$ replaced with $u$. It thus loses its physical integrity, and the two 
oscillators are ``merged'' into a single one that can no longer be decomposed.
Such a control technique is rarely viable for real-world physical or engineered
systems; for instance, replacing one variable in a laser 
device is difficult. Due to that, throughout this paper, we prefer to apply a control law in 
a common way by adding the actuating signal $u$ to one of the derivatives of the 
controlled system \cite{Per05}.

One of the easiest techniques to test for generalized synchronization is the 
auxiliary system procedure, which considers a replica of the response system 
driven by the same driving system. According to this approach, generalized 
synchronization occurs when the response system and its replica are completely 
synchronized for any initial condition \cite{Aba96}. If ${\cal A}_{\rm r}'$ is 
the invariant set produced by the replica of the response system,  
generalized synchronization is such that
${\cal A}_{\rm r} =  {\cal A}_{\rm r}'$. We will restrict ourselves to the 
cases in which generalized synchronization is detected using this technique.
When the drive and response systems are structurally identical, implying 
$d_{\rm d} = d_{\rm r}$ but with a parameter mismatch, it may be 
challenging to clarify the nature of $G$ completely. In that case, the 
departure of the map from the identity is related to the parameter mismatch and 
the coupling strength, such that a threshold value distinguishes generalized 
from complete synchronization \cite{Gut13,Kat13,Min18}.

Pyragas distinguished between two types of generalized synchronization 
when the map $G$ is discontinuous and continuous, respectively \cite{Pyr96}.
When $G$ is continuous, for any $\epsilon \in \mathbb{N}^{+*}$, there exists
$\delta \in \mathbb{N}^{+*}$ such that for all 
$\mb{x}_{\rm d} \in {\cal A}_{\rm d}$ and
$\mb{x}_{\rm d}' \in {\cal A}_{\rm d}$ with 
$|\mb{x}_{\rm d} - \mb{x}_{\rm d}'| < \delta$, then 
$|\mb{x}_{\rm r} - \mb{x}_{\rm r}'| < \epsilon$; conversely, in the 
discontinuous case, $\mb{x}_{\rm r}$ and $\mb{x}_{\rm r}'$ can be arbitrarily
distant from each other. Parlitz also showed that the map $G$ can be mono- or 
multi-valued \cite{Par12}. When $G$ is mono-valued, 
$\mb{x}_{\rm r} \neq \mb{x}_{\rm r}'$ implies that 
$\mb{x}_{\rm d} \neq \mb{x}_{\rm d}'$. Conversely, when $G$ is multivalued, 
there exist $\mb{x}_{\rm r} \neq \mb{x}_{\rm r}'$ such that
$\mb{x}_{\rm d} = \mb{x}_{\rm d}'$. Due to the 
various types of response that can be observed, it is therefore important to 
describe the different types of generalized synchronization which can be 
obtained when the coupling strength is varied.

Among the various types of control law which can be used, we will focus on the
flat ones since it was recently shown that generalized synchronization was 
easily obtained with it over a vast range of coupling strengths \cite{Let23c}. 
Fliess and coworkers introduced flat control law \cite{Fli95}, as being 
based on the constraint related to the algebraic structure of the response 
system with the requirement to provide a measured variable with a global 
observability of the state space: there is thus a diffeomorphism between the 
original state space $\mb{x}
\in \mathbb{R}^{d_{\rm r}}$ and the $s_{\rm d}$-induced $d_{\rm r}$-dimensional 
differential embedding \cite{Let05a}. The actuating signal is applied to the 
derivative of the response system which is the dual of the measured variable
in the sense of Kalman \cite{Kal60} and Lin \cite{Lin74}: there is thus global
controllability. It is furthermore needed that the actuating signal $u$ 
solely appears in the $d_{\rm r}$-th Lie derivative of the measured variable: 
the controlled response system can thus be rewritten in a canonical form 
\cite{Isi81,Isi86,Fli95}. It can be shown that flat control laws are 
efficient for controlling chaotic systems to any desired state \cite{Let21e}. 
Recently, in Ref. \cite{Let23c}, a state of generalized synchronization was 
guaranteed by employing a flat control law based on feedback linearization at 
least within a specific range of coupling strengths. To test the flexibility 
of flat control law toward providing generalized synchronization, we will 
explore extreme cases, such as when $d_{\rm d} \neq d_{\rm r}$ as well as 
when one of the two systems is dissipative and the other is conservative.

By definition, when two structurally different systems are synchronized to each 
other --- even with a straight line as a Lissajous curve between the measured
variables --- they are necessarily in generalized synchronization since the 
other variables are not related in the same way to the measured ones. Here, we 
qualify such generalized synchronization as being strong. The other case, when 
the Lissajous curves between the measured variables used for coupling the two 
systems do not define a straight line --- or any other simple relationship --- 
but rather a complex shape although with an invariant structure, would be 
consequently qualified as representing a weak form of generalized 
synchronization \cite{Pyr96}, with one drive state $\mb{x}_{\rm d}$ leading to 
various response states $\mb{x}_{\rm r}$ \cite{Par12}. This is directly related 
to the coupling strength.

When the coupling strength is decreased, the Lissajous curve may switch from a
straight line to a more complex portrait. Let us define the mean thickness of
the Lissajous curve as
\begin{equation}
  \overline{\epsilon}_{\rm L} =
  \left\langle
    \displaystyle
    \frac{| s_{\rm r} - s_{\rm d}|}{s_{\rm d}^{\rm max} - s_{\rm d}^{\rm min}}
  \right\rangle \,  
\end{equation}
where $s_{\rm d}$ ($s_{\rm r}$) is the variable measured in the response
(drive) system, $s_{\rm d}^{\rm max}$ ($s_{\rm d}^{\rm min}$) is the maximum
(minimum) of the variable $s_{\rm d}$. By analogy with Student's test 
\cite{Stu08} for which the calculated $p$-value must be below a threshold 
(commonly 0.05) chosen for statistical significance, we will consider that this 
is a straight line when $\overline{\epsilon}_{\rm L} < 0.05$ and that, in such 
case, there is a strong generalized synchronization. Then, our guidelines for 
distinguishing the different types of generalized synchronization is provided by 
the response first-return map for a very dissipative three-dimensional drive 
system. As we will show in Section \ref{Sys}, the strongest form of generalized 
synchronization is provided when the 
response first-return map cannot be distinguished from the drive one, for 
instance, it may be shown that the same population of periodic orbits can be 
extracted in both maps: the two maps are topologically conjugated. Furthermore,
we have ${\cal D}_{\rm r} = d_{\rm d}$. When the coupling strength is decreased,
the response map may lose its one-dimensional structure and become thick; we 
will say that the map are $\epsilon$-conjugated meaning that the topological
properties are nearly preserved or, in other words, light discrepancies between
the drive and the response dynamics can be observed (see examples in Section
\ref{Sys}). Typically, for smaller coupling strength, the response first-return 
map becomes thicker or strongly foliated ($\overline{\epsilon}_{\rm L} > 0.05$)
and we can no longer consider the drive and the response maps (or sections) 
as being conjugated, the map $G$ is multivalued or, worst, discontinuous: the 
embedding dimension of the response dynamics is necessarily such that 
${\cal D}_{\rm r} > d_{\rm d}$.

It was shown that different chaotic 
systems coupled with a flat control law exhibit particular forms of generalized 
synchronization  depending on whether the first-return maps of the drive and 
response trajectories were topologically conjugate \cite{Let23c}. In that case, 
there is a homeomorphism relating the two maps \cite{Alli96}, and the response 
trajectory can be embedded within a space whose dimension is at least equal to 
the drive's dimension. As it will be shown later, this property defines two 
other different types of generalized synchronization. There are, therefore, at 
least three ingredients for characterizing generalized synchronization: i) the 
strength of the coupling or control law acting between the two systems, ii) the 
dimension of the space in which the response dynamics takes place, and iii) the 
possible conjugacy between the first-return maps of the two systems. Based on 
them, one can tentatively establish a taxonomy of generalized synchronization 
as reported in Table \ref{typeGS}.

\begin{table}[ht]
  \centering
  \caption{The different types of generalized synchronization  
observed between two systems unidirectionally coupled: they are ranked from 
the strongest to the weakest form of generalized synchronization. $G$ is the map 
between the drive invariant set ${\cal A}_{\rm d}$ and the response invariant 
set ${\cal A}_{\rm r}$. ${\cal D}_{\rm r}$ means the embedding dimension 
$d_{\rm r}$ of the response dynamics.  }
  \label{typeGS}
 \begin{tabular}{cccccc}
	  \\[-0.3cm]
    \hline \hline
	  \\[-0.3cm]
	  Type & $\overline{\epsilon}_{\rm L}$ & ~~~~ ${\cal D}_{\rm r}$ ~~~~
	  & First-return maps & $G$ \\[0.1cm]
    \hline
	  \\[-0.3cm]
    {\sc i}   & $\leqslant 0.05$ & $= d_{\rm d}$ & conjugated & monovalued \\
    {\sc ii}  & $\leqslant 0.05$ & $= d_{\rm d}$ & $\epsilon$-conjugated & 
	 $\approx$ monovalued \\
    {\sc iii} & $> 0.05$ & $> d_{\rm d}$ & non-conjugated & multivalued \\
    {\sc iv}  & $> 0.05$ & $> d_{\rm d}$ & cloudy & discontinuous \\[0.1cm]
    \hline \hline
  \end{tabular}
\end{table}

To the best of our knowledge, the literature devoted to generalized 
synchronization only considers coupling between dissipative systems of the same 
dimension. We will, therefore, also investigate the combination of dissipative 
and conservative systems whose dimensions are not necessarily identical. The 
remainder of this paper is organized as follows. Section \ref{Sys} 
briefly introduces the chaotic systems considered throughout this work and 
mainly describes the feedback linearization designed to build the actuating 
signal imposed on those systems when they act as a response system. Section 
\ref{variousGS} is the central part of this paper and it details various 
combinations of systems using a flat coupling, and shows the generality of
our procedure independently of their dimension and  dissipation rate. Section 
\ref{conc} provides some concluding remarks.

\section{The systems} 
\label{Sys}

In the following, several chaotic systems of different dimensions and 
dissipativeness is introduced, focusing on their observability and 
controllability properties to determine the optimal placement of sensors and 
actuators to design a directed flat coupling between pairs of them using 
feedback linearization. Let 
\begin{equation}
  \label{eq:dri-res}
  \left\{
    \begin{array}{cl}
      \dot {\bm{x}}_1 = &\bm{f}_1 (\bm{x}_1)\\
      \dot{\bm{x}}_2 =& \bm{f}_2 (\bm{x}_2) + u
	    (\xi_1,\xi_2)\cdot \bm{g}
    \end{array}
  \right.
\end{equation}
be a pair of dynamical systems
where $\bm{f}_1$ and $ \bm{f}_2$ govern the different chaotic dynamics of the 
drive and response systems, $u(\xi_1,\xi_2)$ is a scalar actuating 
signal function of the measured variables $\xi_1=h_1(\bm{x}_1) \in \mathbb{R}$ 
and $\xi_2=h_2(\bm{x}_2) \in \mathbb{R}$ for observing the states of drive and 
response systems, respectively, and $\bm{g}$ is the constant vector which 
determines to which derivative $\dot{v}$ the actuating signal $u$ is applied. 

All the systems are integrated using a fourth-order Runger-Kutta scheme
with a time step d$t$. The initial conditions reported for each system are 
thus also used when the systems are coupled.

\subsection{The R\"ossler system}

\begin{figure}[ht]
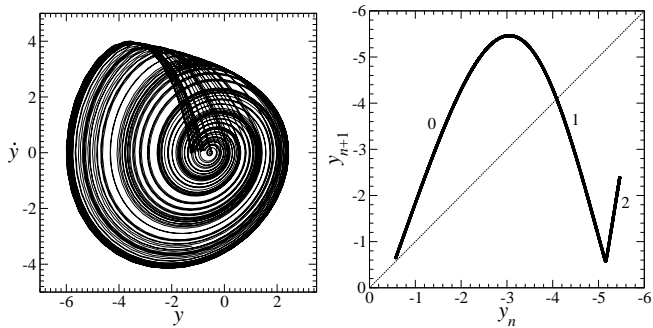

  \centering
  \begin{tabular}{cc}
    \includegraphics[width=0.23\textwidth]{ros452.eps} &
    \includegraphics[width=0.24\textwidth]{rosmap.eps} \\
          (a) State portrait & (b) First-return map \\[-0.3cm]
  \end{tabular}
  \caption{Chaotic attractor produced by the R\"ossler system (\ref{Ros76}). 
The state portrait (a) is spanned by the variable $y_{\rm R}$ and its derivative 
$\dot{y}_{\rm R}$. Parameter values: $a = 0.452$, $b = 2$, and $c = 4$. 
	Initial conditions: $x(0) = y(0) = z(0) = 0.1$ and d$t = 0.01$.}
  \label{normros}
\end{figure}

The R\"ossler system \cite{Ros76c}
\begin{equation}
  \label{Ros76}
  \left\{ 
    \begin{array}{l}
	    \dot{x}  =  - y -z \\[0.1cm]
	    \dot{y}  =  x + a y \\[0.1cm]
	    \dot{z}  =  b - cz + x z
    \end{array} 
  \right.
\end{equation}
is based on a two-dimensional oscillator combined to a switch mechanism
\cite{Ros76a}. It is one of the simplest nonlinear systems producing the most 
simple type of chaotic attractor, that is, characterized by an unimodal folded 
map and without any global torsion \cite{Let95a,Let22a}. Here, instead, we will 
use the R\"ossler system producing a phase non-coherent chaotic attractor 
[Fig.\ \ref{normros}(a)] characterized by a bimodal map [Fig.\ 
\ref{normros}(b)] which has thus three monotone branches. The $d=3$ dimensional 
vector field associated with the R\"ossler system is designated as 
$\bm{f}_{\rm R}(x,y,z)$.

\begin{figure}[ht]
  \centering
  \includegraphics[width=0.43\textwidth]{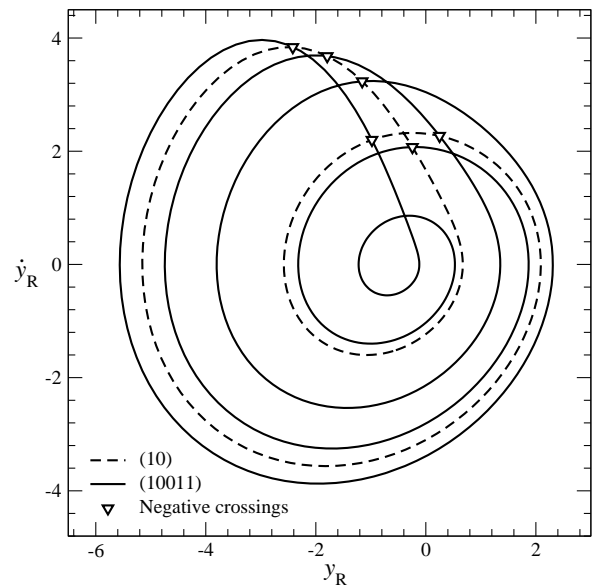} \\[-0.3cm]
  \caption{A knot made of the two periodic orbits (10) and (10011) extracted
from the chaotic attractor produced by the R\"ossler system (\ref{Ros76}). Same 
parameter values as in Fig.\ \ref{normros}.}
  \label{ros_10_10011}
\end{figure}

To compute the first-return map, we use the Poincar\'e section defined as
\begin{equation}
  {\cal P}_{\rm R} \equiv 
  \left\{ \displaystyle
     (x_n, z_n) \in \mathbb{R}^2 ~|~ \dot{y}_n = 0, \ddot{y}_n < 0 
  \right\} \, . 
\end{equation}
Symbolic dynamics is introduced by labeling the left  branch with the even 
symbol ``0'', the middle branch with the odd symbol ``1'', and the right branch 
with the even symbol ``2''. This symbolic dynamics helps encode each 
periodic orbit with a specific symbolic sequence \cite{Bai89}. An example of
two unstable periodic orbits extracted from the R\"ossler attractor by using a
close return technique \cite{Let95a} is plotted in Fig.\ \ref{ros_10_10011}.
To characterize the relative organization of periodic orbits, the linking
number $L_{\rm k} (o_1, o_2)$ between the periodic orbits $o_1$ and $o_2$, is 
computed. The linking number is a topological invariant defined as the half-sum 
of oriented crossings in a regular plane projection of them 
\cite{Tuf92,Let95a}. In the case of the orbits (10) and (10011) [see Fig.\ 
\ref{ros_10_10011} where six negative crossings are counted], the linking 
number is $L_{\rm k} (10,10011) = \frac{-6}{2} = -3$, meaning that the orbit 
(10) rotates three times in the negative direction around the orbit (10011), 
and {\it vice versa}. The positive rotation is associated with a clockwise 
rotation along the flow \cite{Tuf92,Let95a}. The codes for extracting periodic 
orbits from a first-return map, and for computing the linking numbers, can be 
found at the {\sc Atomosyd} website \cite{Ato}.

To quantify the dissipation rate, we used the normalized trace of the Jacobian
matrix
\begin{equation}
	\overline{{\rm Tr} ({\cal J}_{\rm R})} = 
	\frac{1}{T} \, 
	\int{\displaystyle {\rm Tr} ({\cal J}_{\rm R}) \, {\rm d}t}
\end{equation}
where ${\rm Tr} ({\cal J}_{\rm R}) = a - c + x$, and $T$ is the duration
of the simulation. The rate of dissipation does not depend on the timescale but 
only on the topological properties. For the parameter values retained, the 
dissipation rate is $\overline{{\rm Tr} ({\cal J})} = -3.09$.

\begin{figure}[ht]
  \centering
   \begin{tabular}{ccc}
	  \includegraphics[width=0.2\textwidth]{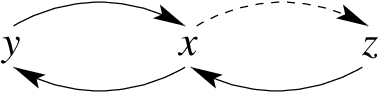} & & 
    \includegraphics[width=0.21\textwidth]{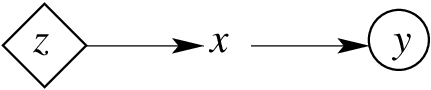} \\
	  (a) Fluence graph & ~~~ & (b) Observability path \\[-0.2cm]
  \end{tabular}
  \caption{Fluence graph (a) and observability path (b) for the R\"ossler
system. The $\circ$ ($\diamond$) designates the placement of the sensor
(actuator).}
  \label{fluparos}
\end{figure}

The observability of the R\"ossler system has already been repeatedly investigated 
(see Ref.  \cite{Let18} and references therein). It is based on the fluence 
graph [Fig.\ \ref{fluparos}(a)] drawn from the symbolic Jacobian matrix 
\begin{equation}
  {\cal J}_{\rm R} = 
  \left[
    \begin{array}{ccc}
      0 & 1 & 1 \\[0.1cm]
      1 & 1 & 0 \\[0.1cm]
      \1 & 0 & \1
    \end{array}
  \right]
\end{equation}
where $1$ ($\1$) stands for constant (non-constant) elements $J_{ji}$
represented as solid (dashed) arrow lines from the $i$th node to the $j$th node 
in the fluence graph \cite{Let18}. For the observability analysis, only the 
constant (linear) elements $J_{\rm ij}$ are considered , therefore, the fluence 
graph is pruned, leaving only the solid line arrows connecting the flow between 
variables \cite{Let18b}. An {\it observability path} is defined as the path 
made of $(d-1)$ linear links visiting all the variables (Fig.\ \ref{fluparos}) 
\cite{Let21e}. The last visited variable is the one to be measured, that is, 
the one at which 
the sensor must be placed to have good observability. The dual variable is the 
one offering good controlability in the sense of Kalman \cite{Kal60} and Lin 
\cite{Lin74}, and is the derivative of the variable from which the linear path 
originates. For the R\"ossler system, the dual of the sensor variable $y$ is 
thus the derivative $\dot{z}$, and this is where the actuator must be placed.

From the analysis of observability and controllability, we have thus 
$h(\mb{x}) = y$ and $\bm{g} = \left[ \begin{array}{ccc} 0 & 0 & 1 \end{array} 
\right]^{\rm T}$. For getting a flat system, this placement must lead to 
${\cal L}_{\bm{g}} {\cal L}_{\bm{f}_{\rm R}}^k h(\mb{x}) = 0$ for 
$0 \leqslant k \leqslant d -2$ and
${\cal L}_{\bm{g}} {\cal L}_{\bm{f}_{\rm R}}^{d-1} h(\bm{x}) \neq 0$, where
${\cal L}_{\bm{f}_{\rm R}}$ is the Lie derivative with respect to the vector 
field $\bm{f}_{\rm R}$ of the R\"ossler system, and ${\cal L}_{\bm{g}}$ is the 
Lie derivative with respect to the input vector field. For the R\"ossler system
with optimal sensor and actuator optimal placement, the flatness condition is
\cite{Fli95}
\begin{equation}
  \label{flatcond}
  \left\{
  \begin{array}{ll}
          h(\mb{x}) = y & {\cal L}_{\bm{g}} \, h(\mb{x}) = 0 \\[0.3cm]
          \displaystyle
          {\cal L}_{\bm{f}_{\rm R}} \, h (\bm{x}) = x + a y
          & {\cal L}_{\bm{g}} \, {\cal L}_{\bm{f}_{\rm R}} \, h(\bm{x}) = 0 \\[0.3cm]
          {\cal L}_{\bm{f}_{\rm R}}^2 \, h (\bm{x}) =
          a x + (a^2 - 1) y - z
          & {\cal L}_{\bm{g}} \, {\cal L}_{\bm{f}_{\rm R}}^2 \, h(\bm{x})
          \displaystyle  = - 1 \neq 0
  \end{array}
  \right.
\end{equation}
Since the first non-zero ${\cal L}_{\bm{g}} \, {\cal L}_{\bm{f}_{\rm R}}^k h(\bm{x})$ is
for
$k = 2 = d -1$, the required condition for a flat control law is fulfilled. It
is therefore possible to design a flat control law with a sensor at $y$ and an
actuator at $\dot{z}$. 

A possible flat control law by feedback linearization is \cite{Isi81}
\begin{equation}
  \label{flatcont}
	  u =
         - \, \frac{
          \kappa_1 (\xi - \xi_{\rm d})
          + \kappa_2 \left( \dot{\xi} - \dot{\xi}_{\rm d} \right)
          + \kappa_3 \left( \ddot{\xi} - \ddot{\xi}_{\rm d} \right)
          + {\cal L}_{\bm{f}}^3 h - \dddot{\xi}_{\rm d}
          }
        {{\cal L}_{\bm{g}} {\cal L}_{\bm{f}}^2 h}
\end{equation}
where $\mb{f} = \mb{f}_{\rm R}$, $\xi=y$, $\xi_{\rm d}$ is the measured 
the variable  from the drive system,
\begin{equation}
        {\cal L}_{\bm{f}_{\rm R}}^3 h = - b + (a^2 - 1) x + a (a^2 -2) y  
        + (c - a) z - xz\, ,
\end{equation}
and
\begin{equation}
        {\cal L}_{\bm{g}} \, {\cal L}_{\bm{f}_{\rm R}}^2 \, h = - 1 \, .
\end{equation}

The $\kappa_i$ ($i \in\{1,\,2, \,3\}$) are
chosen as \cite{Fra15}
\begin{equation}
  \label{kaps}
  \left\{
    \begin{array}{l}
      \kappa_1 =  - \Lambda_1 \Lambda_2\Lambda_3 \\
      \kappa_2 =  \Lambda_1 \Lambda_2 + \Lambda_2 \Lambda_3
            + \Lambda_3 \Lambda_1 \\
      \kappa_3 =  - (\Lambda_1 + \Lambda_2 + \Lambda_3) \, ,
    \end{array}
  \right.
\end{equation}
where $\Lambda_i$ ($i \in\{1,\,2, \,3\}$) are the eigenvalues of the matrix
\[
  \begin{array}{rl}
          A + BK & = \displaystyle
  \left[
    \begin{array}{ccc}
          0 & 1 & 0 \\
          0 & 0 & 1 \\
          0 & 0 & 0
    \end{array}
  \right] \, +
  \left[
    \begin{array}{ccc}
            0 \\ 0 \\ 1
    \end{array}
  \right] \,
  \left[
    \begin{array}{ccc}
            -\kappa_1 & -\kappa_2 & -\kappa_3
    \end{array}
          \right] \\[0.6cm]
          & \displaystyle =
  \left[
    \begin{array}{ccc}
          0 & 1 & 0 \\
          0 & 0 & 1 \\
          -\kappa_1 & -\kappa_2 & -\kappa_3
    \end{array}
  \right] \, ,
  \end{array}
        \]
where $A$ is the Jacobian matrix of the linear part of the canonical form,
$B$ is the matrix defining on which derivative the control law is applied (the 
last one in the case of a feedback linearization), and $K$ is the vector of 
gains used to stabilize the controlled system to the drive dynamics. The 
eigenvalues are chosen to ensure the stability of the closed-loop system. In 
this work, we will use $\Lambda_i = \Lambda$.

\subsection{The FitzHugh-Nagumo system}

In 1948, Bonh\"offer generalized the van der Pol equation \cite{vdP26} to 
obtain a model with one variable for excitability and one for refractoriness
serving as a model for an axon \cite{Bon48}. FitzHugh then used it as a reduced 
model for the Hodgkin-Huxley model of the squid giant axon \cite{Fit61}. It 
reads as
\begin{equation}
  \label{FHNeq}
  \left\{
    \begin{array}{l}
      \displaystyle
	    \dot{x} = \frac{\iota + x -y - \dfrac{x^3}{3}}{\omega} \\[0.3cm]
      \displaystyle
	    \dot{y} = \frac{\alpha + x - \beta y}{\mu \omega} 
    \end{array}
  \right.
\end{equation}
Since FitzHugh's paper, this model is often designated as the Bonh\"offer-van 
der Pol model or the FitzHugh-Nagumo (FHN) model, after Nagumo's work 
\cite{Nag62}. Its Jacobian matrix reads as
\begin{equation}
  J = 
  \left[
    \begin{array}{cc}
      \dfrac{1 - x^2}{\omega} & - \dfrac{1}{\omega} \\[0.3cm]
	    \dfrac{1}{\mu \omega} & - \dfrac{\beta}{\mu \omega}
    \end{array}
  \right] \, . 
\end{equation}
We performed our numerical simulations with the parameter values as follows.
$\alpha = 0.7$, $\beta = 0.8$, $\mu = 12.5$, $\iota = 0.5$, and $\omega = 0.2$:
with those, the FHN system (\ref{FHNeq}) produces a limit cycle (not shown).
The trace of its Jacobian matrix is Tr$({\cal J}_{\rm FHN}) 
= \frac{1 - \beta}{\omega} - \frac{x^2}{\mu \omega}$; the way we normalized its 
mean value is such that it is independent of the $\omega$-value used to tune 
the timescale of the FHN system. For the retained parameter 
values, we got $\overline{{\rm Tr}({\cal J}_{\rm FHN})} = -0.24$. The 
The FHN system is about ten times less dissipative than the R\"ossler system, 
which relaxes more slowly toward a given manifold.

The fluence graph leads to two possible observability paths: 
$x \rightarrow y$ and $y \rightarrow x$. Each variable, therefore, provides a 
global observability and can be selected as the measured variable. We choose to
place the sensor at the variable $x$ and the actuator on the derivative 
$\dot{y}$. In such a case, the flatness conditions (\ref{flatcond}) reads
\begin{equation}
  \left\{
  \begin{array}{ll}
  h(\mb{x}) = x & {\cal L}_{\bm{g}} \, h(\mb{x}) = 0 \\[0.3cm]
          \displaystyle
	  {\cal L}_{\bm{f}} \, h (\bm{x}) 
	  = \frac{\iota + x - y - \frac{x^3}{3} }{\omega} & 
	  {\cal L}_{\bm{g}} \, {\cal L}_{\bm{f}} \, 
	  h(\bm{x}) = -1 \neq 0 
  \end{array}
  \right.
\end{equation}
and is fulfilled. The two variables of the FHN system can be expressed as the 
Lie derivative as
\begin{equation}
  \left|
    \begin{array}{l}
	      x = X \\
	      \displaystyle 
	      y = \iota + X - Y - \frac{X^3}{3} 
    \end{array}
  \right.
\end{equation}
where $X = x$ and $Y = \dot{x}$, thus allowing rewriting the FHN system in 
canonical form as
\begin{equation}
  \left\{
    \begin{array}{l}
       \dot{X} = Y \\[0.1cm]
	    \displaystyle
	    \dot{Y} = \frac{\beta \iota - \alpha + (\beta - 1) X 
	    + \omega (\mu - \beta) Y - \frac{\beta}{3} X^3 - \mu \omega X^2Y}
	    {\mu \omega}
    \end{array}
  \right.
\end{equation}

When feedback linearization is used as the control law, we get the actuating
signal
\begin{equation}
  u_{\rm FHN} = 
         - \, \frac{
          \kappa_1 (\xi - \xi_{\rm d})
          + \kappa_2 \left( \dot{\xi} - \dot{\xi}_{\rm d} \right)
          + {\cal L}_{\bm{f}}^2 h - \ddot{\xi}_{\rm d}
          }
        {{\cal L}_{\bm{g}} {\cal L}_{\bm{f}} h} \, , 
\end{equation}
where 
\begin{equation}
	{\cal L}_{f}^2 h =
	\frac{b \iota -a}{\omega\2 \mu} 
	+ \frac{(b - 1) x}{\mu \omega^2} 
	+ \left(1 - \frac{b}{\mu} \right) \frac{y}{\omega} 
	- \frac{x^2 y}{\omega} 
	- \frac{b x^3}{3 \omega^2 \mu} \, , 
\end{equation}
${\cal L}_{\bm{g}} {\cal L}_{\bm{f}} h = - 1$, and $\kappa_i$s obtained 
as above.

\subsection{The Saito system}

To investigate hyperchaos, Toshimichi Saito used an electronic
circuit made of a linear negative conductor, a three-segment piecewise-linear 
resistor, one small inductor $L_0$ serially connected with it, two capacitors, 
and a second inductor \cite{Sai90}. The ordinary differential equations 
governing the dynamics of this circuit are
\begin{equation}
  \label{saiteq}
  \left\{
    \begin{array}{rl}
            \dot{x} & = \rho (-y +z) \\[0.1cm]
            \dot{y} & = x + 2 \delta y \\[0.1cm]
            \dot{z} & = -x -w  \\[0.1cm]
            \epsilon \dot{w} & = z - \Phi (w)
    \end{array}
  \right.
\end{equation}
where $\Phi$ is the piecewise-linear function
\begin{equation}
  \Phi (w) = 
  \left|
    \begin{array}{lcl}
      w - (1 + \eta) & & w \geq \eta \\[0.1cm]
      \displaystyle
            - \frac{w}{\eta} & \mbox{ for } & |w| < \eta \\[0.1cm]
      w + (1 + \eta) & & w \leq -\eta  \, , 
    \end{array}
  \right.
\end{equation}
and the $d=4$ dimensional vector flow is denoted by $\bm{f}_{\rm S}(x,y,z,w)$. 
When $\delta$ is varied, the dynamics can be chaotic, quasi-periodic, toroidal 
chaotic, or even hyperchaotic, also structured around a torus \cite{Sai90}. 
Since the state space is four-dimensional, the topology of the observed
attractors cannot be topologically characterized. This system is equivariant 
under an inversion symmetry, that is,
\begin{equation}
  \bm{f}_{\rm S} (\Gamma \cdot \mb{x}) = \Gamma \cdot \bm{f}_{\rm S} (\mb{x}) 
\end{equation}
where $\Gamma = - \mathbb{I}$ and $\mathbb{I}$ is the identity
matrix \cite{Let01}.

\begin{figure}[ht]
  \centering
  \includegraphics[width=0.48\textwidth]{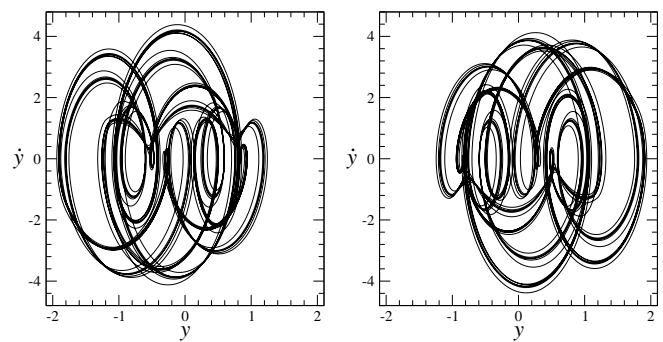} \\
  (a) Plane projections of the state portrait \\[0.2cm]
  \includegraphics[width=0.30\textwidth]{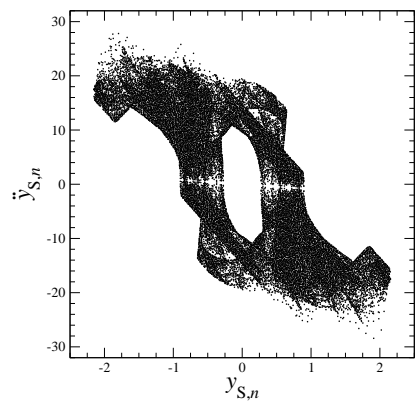} \\
  (b) Poincar\'e section ${\cal P}_{\rm S}$ \\[-0.2cm]
  \caption{Toroidal chaos produced by the Saito circuit (\ref{saiteq}).
The initial conditions are mapped into $(-x_0,-y_0,-z_0,-w_0)$ in the right
panel to exhibit the inversion symmetry of the circuit. Parameter values:
$\eta = 1$, $\rho = 14.4$, $\delta = 0.785$, and $\epsilon = 0.009$. Initial
conditions: $x(0) = 3.908 \cdot 10^{-14}$, $y(0) = 0.0009854$, 
$z(0) = 0.041631$, $w(0) = 0.176643$ and d$t = 0.001$.}
  \label{saitodyn}
\end{figure}

We set the parameter values giving rise to a hyperchaotic regime [Fig.\ 
\ref{saitodyn}(a)] which is close to a limit cycle in the bifurcation diagram 
(not shown). The corresponding Poincar\'e section 
\begin{equation}
  {\cal P}_{\rm S} \equiv
  \left\{ \displaystyle
	(y_n, \ddot{y}_n, \dddot{y}_n) \in \mathbb{R}^3 ~|~
	\dot{y}_n = 0
  \right\}
\end{equation}
shows that the attractor has a toroidal structure [Fig.\ \ref{saitodyn}(b)]; 
its thickness reveals a complex foliated structure, which is necessarily due to 
the hyperchaotic nature of the dynamics since the system is strongly 
dissipative. 

The Jacobian matrix of the Saito system (\ref{saiteq}) is
\begin{equation}
  \label{saitojac}
  {\cal J}_{\rm S} =
  \left[
    \begin{matrix}
            0 & -\rho & -\rho & 0 \\[0.1cm]
            1 & 2 \delta & 0 & 0 \\[0.1cm]
            -1 & 0 & 0 & -1 \\[0.1cm]
            0 & 0 & \displaystyle \frac{1}{\epsilon} & 
            \displaystyle - \frac{\Phi_w}{\epsilon}
    \end{matrix}
  \right] \, .
\end{equation}
where
\begin{equation}
  \Phi_w = \frac{\partial \Phi}{\partial w}  
  = \left|
   \begin{array}{rl}
           \displaystyle - \frac{1}{\eta} & \mbox{ if } |w| < \eta \\[0.3cm]
           \displaystyle 1 & \mbox{ if } |w| \geqslant \eta  \, .
   \end{array}
  \right.
\end{equation}
The element ${\cal J}_{44}$ is the only term depending on the location in the 
state space $\mathbb{R}^4 (\mb{x})$. The trace is
\begin{equation}
  \mbox{Tr} ({\cal J}_{\rm S}) = 2 \delta +
  \left|
    \begin{array}{ll}
	    \displaystyle 
	    + \frac{1}{\eta \epsilon} & \mbox{ if } |w| \le \eta \\[0.3cm]
	    \displaystyle 
	    - \frac{1}{\epsilon} & \mbox{ otherwise.}
    \end{array}
  \right.
\end{equation}
The system is therefore expansive for $|w| \leqslant \eta$ and positive values
of the parameters; for $|w| > \eta$, it is dissipative when 
$\delta < \frac{1}{2 \epsilon}$. With the retained parameter values, 
$\overline{{\rm Tr} ({\cal J}_{\rm S})} = - 106$: this system is therefore
strongly dissipative (30 times more than the R\"ossler system).

\begin{figure}[ht]
  \centering
  \includegraphics[width=0.26\textwidth]{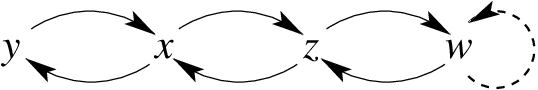} \\
        (a) Fluence graph \\[0.2cm]
  \includegraphics[width=0.26\textwidth]{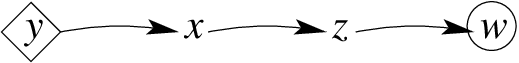} \\
        (b) Observability path 1 \\[0.2cm]
  \includegraphics[width=0.26\textwidth]{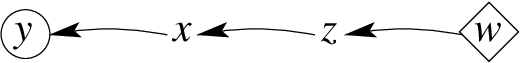} \\
        (c) Observability path 2 \\[-0.2cm]
  \caption{Fluence graph of the Saito system (\ref{saiteq}) and the two
possible observability paths. The sensor is placed in the circled variable, and
the actuator is at the squared variable's derivative.}
  \label{flusaito}
\end{figure}

For the observability analysis, the element ${\cal J}_{44}$ is considered as
being non-constant, leading to the symbolic Jacobian matrix \cite{Let18}
\begin{equation}
  \label{symsaitojac}
  J_{\rm S} =
  \left[
    \begin{matrix}
            0 & 1 & 1 & 0 \\[0.1cm]
            1 & 1 & 0 & 0 \\[0.1cm]
            1 & 0 & 0 & 1 \\[0.1cm]
            0 & 0 & 1 & \1
    \end{matrix}
  \right] \, .
\end{equation}
from which the fluence graph is drawn as in Fig.\ 
\ref{flusaito}(a). Disregarding the dashed arrow line that accounts for the 
nonlinear term, there are two observability paths, as shown in Figs.\ 
\ref{flusaito}(b) and \ref{flusai,to}(c). From the observability point of 
view, measuring a variable whose derivative is directly affected by a nonlinear 
component is often a source of singularities, particularly when a 
piecewise linear function is involved \cite{Let23b}. In the present case, the 
switching mechanism implies that $\ddot{w}$ is not defined in the the planes
\begin{equation}
  {\cal M}^{\rm sw} \equiv
  \left\{
    (x,y,z) \in \mathbb{R}^3 ~|~ w = \pm \eta 
  \right\} \, ;
\end{equation}
consequently, there is a singular observability manifold: the
observability is only local when the variable $w$ is measured. The second
observability path [Fig.\ \ref{flusaito}(c)] is therefore retained, that is, we 
placed the sensor at variable $y$ and the actuator at the derivative $\dot{w}$.

When the Saito system is driven through a flat control law, the actuating 
signal applied to $\dot{w}$ is
\begin{equation}
  \label{actsig4D}
  \begin{array}{rl}
          u_{\rm S} & = - \left[ \displaystyle \kappa_1 (y - \xi_{\rm d})
          + \kappa_2 \left( \dot{y} - \dot{\xi}_{\rm d} \right)
          \right. \\[0.1cm]
          & \frac{\left. \displaystyle ~~~~~~
          + \kappa_3 \left( \ddot{y} - \ddot{\xi}_{\rm d} \right)
          + \kappa_4 (\dddot{y} - \dddot{\xi}_{\rm d})
          + {\cal L}_{\bm{f}_{\rm S}}^4 h - \ddddot{\xi}_{\rm d} 
	  \right]}{\displaystyle
          {\cal L}_{\bm{g}} {\cal L}_{\bm{f}_{\rm S}}^3 h}
  \end{array}
\end{equation}
where $h(\bm{x})=y$,
\begin{widetext}
\begin{equation}
  \label{saitflat-ydw}
    {\cal L}^4_{\bm{f}_{\rm S}} h =
     - \frac{\rho}{\epsilon} \,y
     + 2 \left( \displaystyle \rho \, \delta + \frac{\delta}{\epsilon} \right)
     \dot{y}
     - \left( \displaystyle 2 - \frac{1}{\epsilon} \right) \, \ddot{y}
                  + 2 \delta \, \dddot{y} +
     \left|
       \begin{array}{lcl}
               \displaystyle   \frac{ \displaystyle
                  + 2 \rho \, \delta \, y - 2 \rho \, \dot{y}
                  + 2 \delta \, \ddot{y} - \dddot{y}  + 2 \rho }{\epsilon} & &
                  w \leqslant - \eta \\[0.3cm]
               \displaystyle \frac{ \displaystyle
                  - 2 \rho \, \delta \, y + 2 \rho \, \dot{y}
                  - 2 \delta \, \ddot{y} + \dddot{y} }{\epsilon}
                  &  \mbox{ if } & |w| < \eta \\[0.3cm]
               \displaystyle \frac{ \displaystyle
                  + 2 \rho \, \delta \, y - 2 \rho \, \dot{y}
                  + 2 \delta \, \ddot{y} - \dddot{y}  - 2 \rho }{\epsilon} & &
                  w \geqslant + \eta
       \end{array}
     \right.
\end{equation}
\end{widetext}
and ${\cal L}_{\bm{g}} {\cal L}_{\bm{f}_{\rm S}}^3 h = - \rho$. The $\kappa_i$ values are chosen as in Eq.~(\ref{kaps}).

\subsection{The Sprott A system}

\begin{figure}[ht]
  \centering
  \includegraphics[width=0.45\textwidth]{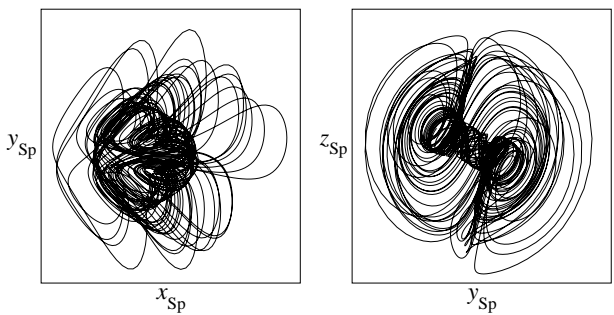} \\
  (a) Plane projections of the state portrait \\[0.2cm]
  \includegraphics[width=0.38\textwidth]{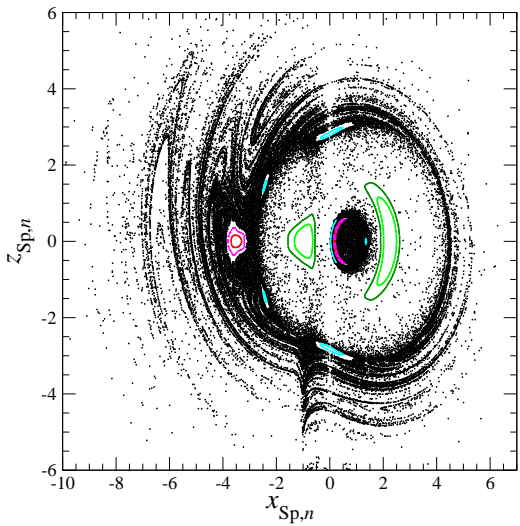} \\
  (b) Poincar\'e section \\[-0.2cm]
  \caption{Dynamics produced by the Sprott A system (\ref{poisecSA}).
Depending on initial conditions, the behavior is characterized by a chaotic sea 
(black) and quasi-periodic islands (colored closed curves). Parameter values: 
$a = 0.452$, $b = 2$, $c = 4$, $\alpha = 1$, and $\beta = 0.65$. Initial
conditions (for the chaotic sea): $x_0 = 3$, $y_0 = 0$, $z_0 = 0$, and 
d$t = 0.02$.}
  \label{sproAsec}
\end{figure}

The Sprott A system reads 
\begin{equation}
  \left\{
    \begin{array}{l}
      \dot{x} = y + z \\[0.1cm]
      \dot{y} = -x + yz \\[0.1cm]
	    \dot{z} = \alpha - x - \beta y^2 \,  
    \end{array}
  \right.
\end{equation}
is a mathematical system obtained by a brute-force search for simple chaotic
systems \cite{Spr94}. The trace Tr(${\cal J}_{\rm Sp}) = z$ and its mean value 
are null: the system is as observed in every conservative system \cite{Hen64}, 
depending on the initial conditions for a given set of parameter values, the 
behavior can be chaotic, quasi-periodic or periodic. There is no attractor but 
an invariant set. This is easily seen with the Poincar\'e section defined as
\begin{equation}
  \label{poisecSA}
	{\cal P}_{\rm Sp} \equiv
	\left\{ 
	\displaystyle
	(x_{n}, z_{n}) \in \mathbb{R}^2
	~|~ y_{n} = 0, \dot{y}_{n} \gtrless 0
	\right\}
\end{equation}
where the chaotic sea surrounds the quasi-periodic islands [Fig.\ 
\ref{sproAsec} where parameter values are reported]. The trace of the 
Jacobian matrix is Tr$({\cal J}_{\rm Sp}) = z$ and its mean value is null.
There is no attractor for this type of system but co-existing invariant sets. 

\begin{figure}[ht]
  \centering
  \begin{tabular}{ccc}
	  \includegraphics[width=0.2\textwidth]{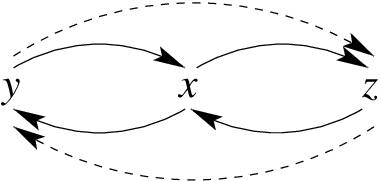} & & 
    \includegraphics[width=0.21\textwidth]{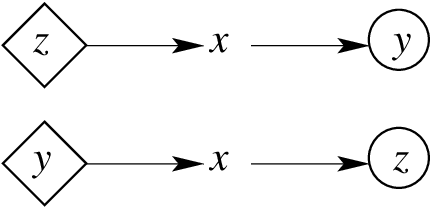} \\
	  (a) Fluence graph & ~~~ & (b) Observability paths \\[-0.2cm]
  \end{tabular}
  \caption{Fluence graph (a) and observability paths (b) for the Sprott A
system. The $\circ$ ($\diamond$) designates the placement of the sensor
(actuator).}
  \label{fluenceSA}
\end{figure}

The symbolic Jacobian matrix of the Sprott A system is
\begin{equation}
  J_{\rm Sp} = 
  \left[
    \begin{array}{ccc}
      0 & 1 & 1 \\[0.1cm]
      1 & \1 & \1 \\[0.1cm]
      1 & \1 & 0
    \end{array}
  \right] \, , 
\end{equation}
and leads to the fluence graph in Fig.\ \ref{fluenceSA}, from which two
observability paths are obtained. Both suggest measuring a variable,
$y$ or $z$, whose derivative is affected by a nonlinear term, a property which,
as said above, it leads to a singular observability manifold. 
There is, therefore, no variable providing a global observability of the state 
space for the Sprott A system. It is thus necessary to examine the determinants of 
the observability matrices, which are
Det$~{\cal O}_{x^3} = (2 \beta + 1) y - z$,
Det$~{\cal O}_{y^3} = 1 + x + y (z -y)$, and 
Det$~{\cal O}_{z^3} = 1 - 2 \beta x + 4 \beta y (y + z)$. Since the 
determinants 
of the observability matrix with the smallest degree is Det$~{\cal O}_{x^3}$, 
the variable $x$ is therefore the best variable to measure \cite{Let02}. By 
duality, there is no derivative providing global controllability. There is no simple possibility to find a placement for a sensor and an actuator 
providing flatness: this directly stems from the fact that every derivative 
is affected by the two other variables, leading to 
${\cal L}_{\bm{g}} \, {\cal L}_{\bm{f}} \, h \neq 0$ for any choice of $h$ and 
$\bm{g}$. Designing a flat control law for this system is, therefore, not a 
trivial task which will not be considered here.

\subsection{The 4D conservative H\'enon system}

\begin{figure}[ht]
  \centering
    \includegraphics[width=0.42\textwidth]{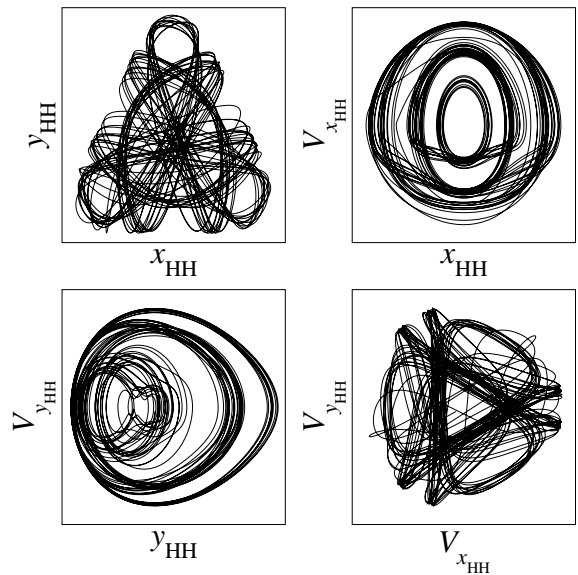} \\
    (a) Plane projections of the state portrait \\[0.2cm]
    \includegraphics[width=0.40\textwidth]{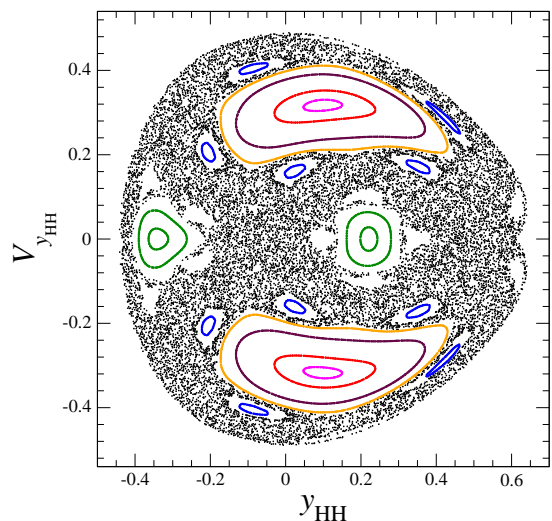} \\
	(b) Poincar\'e section \\[-0.2cm]
  \caption{Dynamics of the conservative H\'enon-Heiles system (\ref{4Dheinon}). 
Parameter values: $E = 0.126607335$. The chaotic sea surrounds the 
quasi-periodic islands. Initial conditions for the chaotic sea: $x_0 = 0$, 
$y_0 = - 0.15$, $V_{x,0} = \sqrt{2E - [1 - \frac{2y(0)}{3}] y^2(0) 
- V_y^2(0)}$, $V_y(0) = 0.25$, and d$t = 0.002$.}
  \label{henhei}
\end{figure}

The four-dimensional H\'enon-Heiles system 
\begin{equation}
  \label{4Dheinon}
  \left\{
    \begin{array}{l}
      \dot{x} = V_x \\[0.1cm]
      \dot{V}_x = -x - 2 xy \\[0.1cm]
      \dot{y} = V_y \\[0.1cm]
      \dot{V}_y = -y + y^2 - x^2 \, 
    \end{array}
  \right.
\end{equation}
is a prototypical system for governing a star's motion in a galaxy's potential \cite{Hen64}. It
is a conservative system whose symbolic Jacobian matrix is
\begin{equation}
  \label{henheijac}
 J_{\rm H} =
  \left[
    \begin{matrix}
            0 & 1 & 0 & 0 \\[0.1cm]
            \1 & 0 & \1 & 0 \\[0.1cm]
            0 & 0 & 0 & 1 \\[0.1cm]
            \1 & 0 & \1 & 0
    \end{matrix}
  \right] \, ,
\end{equation}
and whose trace is null. The state portrait has a toroidal structure, as shown 
in some plane projections of the state portrait [Fig.\ \ref{henhei}(a)]. The
system is equivariant under the rotation defined by the matrix:
\begin{equation}
  \Gamma = 
  \left[
    \begin{array}{cccc}
      - 1 & 0 & 0 & 0 \\
      0 & - 1 & 0 & 0 \\
      0 & 0 & + 1 & 0 \\
      0 & 0 & 0 & + 1 
    \end{array}
  \right] \, , 
\end{equation}
that is, $\mb{f} (\Gamma \cdot \mb{x}) = \Gamma \cdot \mb{f} (\mb{x})$ 
\cite{Let01}. This symmetry can be seen in the symmetry observed in the 
$x$-$y$ and $x$-$V_x$ plane projections in Fig.\ \ref{henhei}(a). Moreover, 
there is an obvious symmetry in the Poincar\'e section [Fig.\ \ref{henhei}(b)]
which results from the fact that time can be reversed in a conservative system 
\cite{Gut90}, thus explaining why the variable $V_{y_{\rm HH}}$ can be 
transformed into its opposite.

\begin{figure}[ht]
  \centering
  \begin{tabular}{ccc}
	  \includegraphics[width=0.11\textwidth]{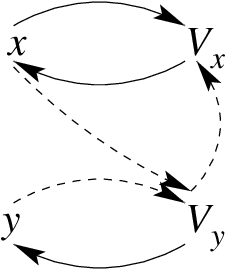} & & 
    \includegraphics[width=0.12\textwidth]{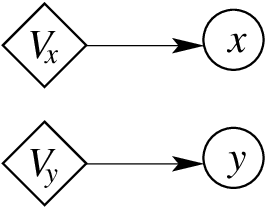} \\
	  (a) Fluence graph & ~~~ & (b) Observability paths \\[-0.2cm]
  \end{tabular}
  \caption{Fluence graph (a) and observability path (b) for the H\'enon-Heiles
system. The $\circ$ ($\diamond$) designates the placement of the sensor
(actuator).}
  \label{fluhenon}
\end{figure}

The reduced fluence graph of the H\'enon-Heiles system 
reveals two disconnected components [Fig.\ \ref{fluhenon}(a)], meaning that 
two variables need to be measured, namely, $x$ and $y$. Indeed, the 
determinant of the observability matrix is Det${\cal O}_{x^2y^2} = 1$. Since it 
is associated with a linear transformation of coordinates, the state space is 
globally observable from these variables \cite{Cha19}.

By duality, since each actuator should correspond to an actuator, two actuators
must be placed on the H\'enon-Heiles system. They should be placed on the 
derivatives $\dot{V}_x$ and $\dot{V}_y$, respectively [Fig.\ 
\ref{fluhenon}(b)]. Two actuating signals must be designed according to
\begin{equation}
  \label{actsighenx}
	  u_x = - \, \frac{ 
          \kappa_1 (x - \xi_{\rm d}) 
	  + \kappa_2 (\dot{x} - \dot{\xi}_{\rm d}) 
	  + {\cal L}_{\bm{f}_{\rm H}}^2 h_x - \ddot{\xi}_{\rm d}
	  }
	    {{\cal L}_{\bm{g} } {\cal L}_{\bm{f}_{\rm H}} h_x}
	    \, ,
\end{equation}
for $h_x (\mb{x}) = x$, and 
\begin{equation}
  \label{actsigheny}
	  u_y = - \, \frac{ 
          \kappa_1 (y - \xi_{\rm d}) 
          + \kappa_2 (\dot{y} - \dot{\xi}_{\rm d}) 
	  + {\cal L}_{\bm{f}_{\rm H}}^2 h_y - \ddot{\xi}_{\rm d}
	  }
	    {{\cal L}_{\bm{g}} {\cal L}_{\bm{f}_{\rm H}} h_y} \, ,
\end{equation}
for $h_y (\mb{x}) = y$. In these equations,  
${\cal L}_{\bm{g}} {\cal L}_{\bm{f}_{\rm H}} h_x = {\cal L}_{\bm{g}} {\cal L}_{\bm{f}_{\rm H}} h_y = 1$ and 
\begin{equation}
  \left\{
    \begin{array}{l}
      \displaystyle {\cal L}_{\bm{f}_{\rm H}}^2 h_x  = - x -2 xy \\[0.1cm]
      \displaystyle {\cal L}_{\bm{f}_{\rm H}}^2 h_y  = - y -x^2 + y^2 .
    \end{array}
  \right.
\end{equation}

\section{Generalized synchronization}
\label{variousGS}

\begin{figure}[ht]
  \centering
  \includegraphics[width=0.48\textwidth]{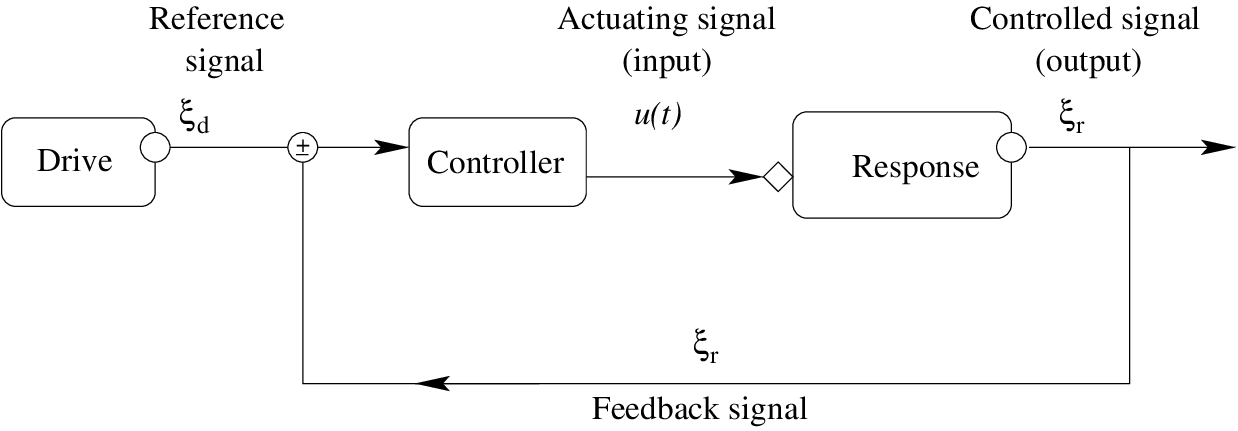} \\[-0.3cm]
  \caption{Block diagram for a flat coupling between a drive 
and response systems. $\meddiamond \equiv$ actuator and 
$\medcircle \equiv$ sensor.}
  \label{block_SR}
\end{figure}

In this section, two different dynamical systems are flat coupled according to
the block diagram drawn in Fig.\ \ref{block_SR}. To do this, one 
variable $\xi_{\rm d} = h_{\rm d} (\mb{x}_{\rm d}$ is measured in the drive
system and one variable $\xi_{\rm r} = h_r (\mb{x}_r)$ in the response system.
The actuating signal $u = u(\xi_{\rm d},\xi_{\rm r})$ is a function of these
two measured variables. When a flat control law is used, the $d_{\rm d}$ and 
$d_{\rm r}$ derivatives of the measured variables $\xi_{\rm d}$ and 
$\xi_{\rm r}$ provide a global observability of the state spaces
$\mathbb{R}^{d_{\rm d}}$ and $\mathbb{R}^{d_{\rm r}}$, respectively. Therefore, there are at least two sensors on each system. In most cases, a single actuator is placed on the derivative of the variable, which is the 
dual of the measured variable $\xi_{\rm r}$. The coupling strength is here 
quantified with the constant $\Lambda$ determining the $\kappa_i$s.

\subsection{R\"ossler driving FitzHugh-Nagumo}

This case is particularly interesting since a two-dimensional system is 
driven by a chaotic one. The fact that the response system is two-dimensional
does not prevent to observe a chaotic response dynamics since it is known that 
a driven two-dimensional oscillator can produce chaos \cite{Ued65,Par85}.

\begin{figure}[ht]
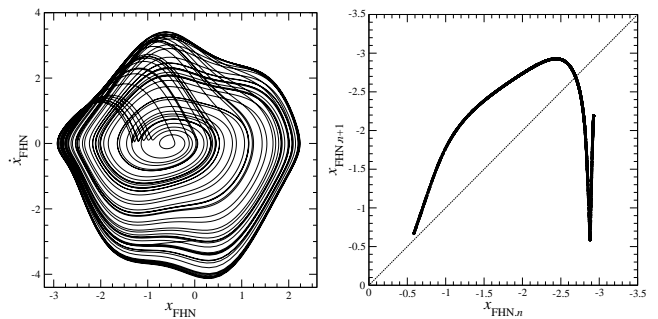
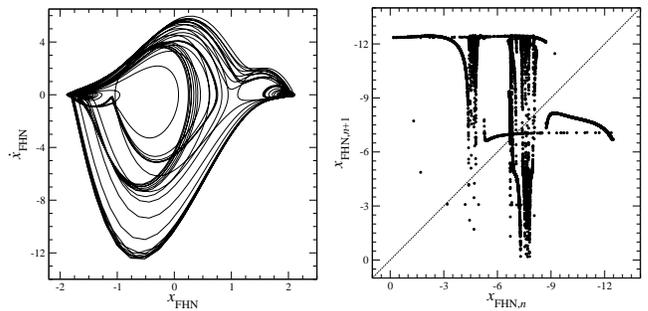

  \centering
  \begin{tabular}{cc}
    \includegraphics[width=0.23\textwidth]{flatFHN_80.eps} &
    \includegraphics[width=0.236\textwidth]{flatFHN_map80.eps} \\
	  \multicolumn{2}{c}{(a) $\Lambda = - 8.0$: 
	  $\overline{\epsilon}_{\rm L} < 0.05$: type-{\sc i} generalized 
	  synchronization} \\[0.2cm]
    \includegraphics[width=0.23\textwidth]{flatFHN_20.eps} &
    \includegraphics[width=0.228\textwidth]{flatFHN_map20.eps} \\
    \multicolumn{2}{c}{(b) $\Lambda = - 2.0$: 
	  $\overline{\epsilon}_{\rm L} < 0.11$: type-{\sc ii} generalized 
	  synchronization} \\[0.2cm]
    \includegraphics[width=0.23\textwidth]{flatFHN_08.eps} &
    \includegraphics[width=0.228\textwidth]{flatFHN_map08.eps} \\
   \multicolumn{2}{c}{(c) $\Lambda = - 0.8$: 
	  $\overline{\epsilon}_{\rm L} < 0.30$: type-{\sc iv} generalized 
	  synchronization} \\[-0.2cm]
  \end{tabular}
  \caption{Response FigtzHugh-Nagumo dynamics to the drive R\"ossler system.}
  \label{FlatFHNRos}
\end{figure}

When the coupling strength is sufficient ($\Lambda = - 8$ and 
$\overline{\epsilon}_{\rm L} < 0.05$), there is a strong generalized 
synchronization, and the response dynamics [Fig.\ \ref{FlatFHNRos}(a)] is 
topologically equivalent to the drive one [Fig.\ \ref{normros}]; in particular, 
the first-return maps are topologically conjugate [compare Fig.\ 
\ref{normros}(b) with the right panel of Fig.\ \ref{FlatFHNRos}(a)]. When the 
coupling strength is reduced ($\Lambda = - 2$ and $\overline{\epsilon}_{\rm L}
= 0.11$), the first-return map of the response dynamics becomes foliated 
[Fig.\ \ref{FlatFHNRos}(b)] and additional structures emerge within the state 
portrait, increasing the embedding dimension ${\cal D}_{\rm r}$. This is 
a type-{\sc ii} generalized synchronization. For a smaller coupling
strength ($\Lambda = - 0.8$ and $\overline{\epsilon}_{\rm L} = 0.30$): the 
generalized synchronization is weak. The response first-return map is not only 
foliated but also discontinuous [Fig.\ \ref{FlatFHNRos}(c)] as a signature of 
discontinuous multivalued $G$. The structure of the drive dynamics as well as 
the structure of the response one is blurred: this is a type-{\sc iv} 
generalized synchronization (see Table \ref{typeGS}).

\subsection{Sprott A driving R\"ossler}

\begin{figure}[ht]
  \centering
    \includegraphics[width=0.38\textwidth]{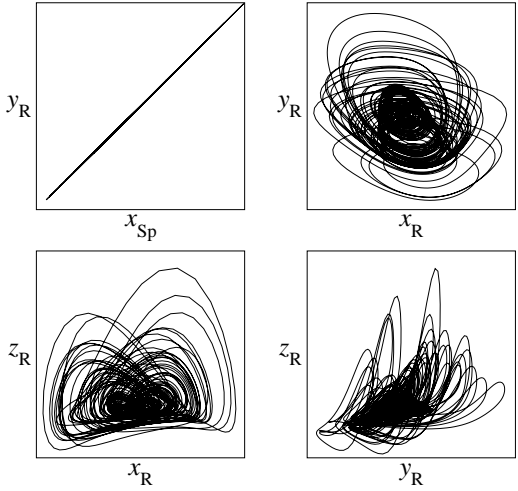} \\
    (a) Plane projection of the state space \\[0.2cm]
    \includegraphics[width=0.38\textwidth]{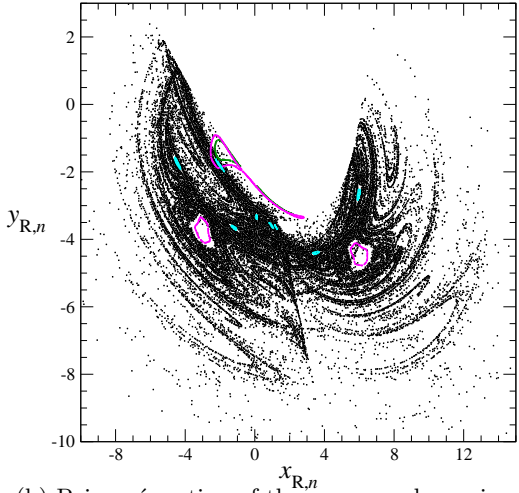} \\[-0.2cm]
    (b) Poincaré section of the response dynamics \\[-0.2cm]
  \caption{Dynamics of the response R\"ossler system (\ref{poisecRres}). 
Initial conditions are varied as in Fig.\ \ref{sproAsec}. Parameter values as 
in Figs.\ \ref{normros} and \ref{sproAsec}. Control parameter value: 
	$\Lambda = - 16$ and $\overline{\epsilon}_{\rm L} < 10^{-4}$.}
  \label{RosproA}
\end{figure}

We now drive the 3D dissipative R\"ossler system with the 3D conservative 
Sprott A system. We did that for the different behaviors observed within the 
Poincar\'e section defined as in Eq.~(\ref{poisecSA}) and plotted in Fig.\ 
\ref{RosproA}. A Poincar\'e section of the response R\"ossler system was 
defined as
\begin{equation}\label{poisecRres}
	{\cal P}_{\rm R} \equiv
	\left\{ 
	\displaystyle
	(x_{{\rm R},n}, y_{{\rm R},n}) \in \mathbb{R}^2
	~|~ z_{{\rm R},n} = 0, \dot{z}_{{\rm R},n} \gtrless 0
	\right\}
\end{equation}
and plotted in Fig.\ \ref{sproAsec}(b). The chaotic sea --- the now common 
name given by H\'enon and Heiles to the set of intersections of a chaotic
trajectory to a Poincar\'e section \cite{Hen64} --- is recovered with some 
quasi-periodic islands, although some of them are difficult to distinguish due
to some self-overlaps of the dynamics. We did not find another Poincar\'e 
section with a better unfolding. This results from the different ways the 
variables are coupled in each system [compare Figs.\ \ref{normros}(a) and
\ref{fluenceSA}(a)]. The Poincar\'e section is distorted with many overlaps: 
moreover, the four ``blue'' quasi-periodic islands in Fig.\  \ref{sproAsec}(b)
are now eight in the response Poincar\'e section [Fig.\ \ref{RosproA}(b)].
The $\Lambda$-value cannot be increased to remove this multivalued response:
this is a type-{\sc iii} generalized synchronization. Since 
$\overline{\epsilon}_{\rm L} < 10^{-4}$, this is a strong generalized 
synchronization. Perhaps the dissipative nature of the R\"ossler system is 
responsible for the overlap of the dynamics. At this point, it is important to 
realize that the coordinate transformation between the Sprott A invariant set 
and the response invariant set is not a static map but a dynamic one, since it 
depends on the derivatives of the R\"ossler system. 

\subsection{H\'enon-Heiles driving R\"ossler}

\begin{figure}[ht]
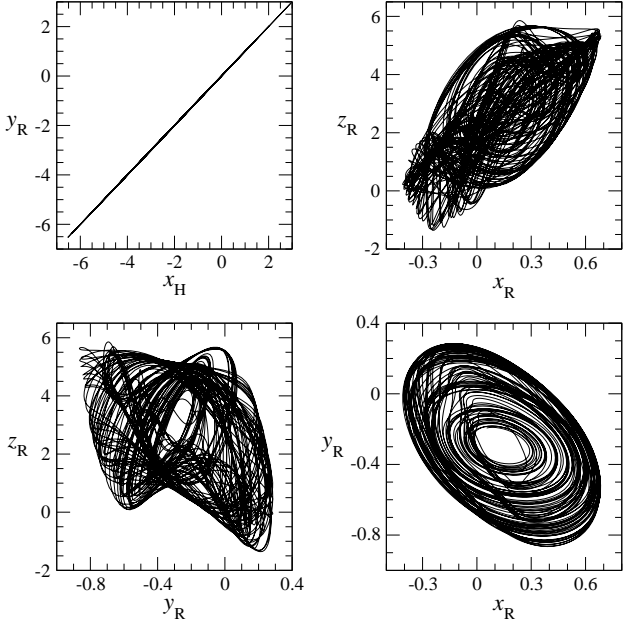

  \centering
  \includegraphics[width=0.46\textwidth]{henoros.eps} \\
	(a) Plane projection of the state space \\[0.2cm]
  \includegraphics[width=0.42\textwidth]{henorosec.eps} \\
	(b) Poincar\'e section of the response dynamics \\[-0.2cm]
  \caption{Dynamics of the response R\"ossler system (\ref{eq:dri-res}) driven 
by the H\'enon-Heiles system (\ref{4Dheinon}). In the Poincar\'e section, the 
chaotic sea (black) surrounds quasi-periodic islands (colors). Control parameter 
value: $\Lambda = - 30$ and $\overline{\epsilon}_{\rm L} < 10^{-4}$.}
  \label{henoros}
\end{figure}

We are here investigating another case for which the four-dimensional 
conservative H\'enon-Heiles system drives the three-dimensional dissipative 
R\"ossler system. To do that, it is helpful to rescale the R\"ossler system to 
have the measured variable $y_{\rm R}$ fluctuating within the same range as 
the variable $x_{\rm H}$ of the H\'enon-Heiles system. The rescaled R\"ossler 
system reads,
\begin{equation}
  \label{rescaros}
  \left\{
    \begin{array}{l}
      \dot{x}_{\rm R} = - y_{\rm R} - \rho z_{\rm R} \\[0.1cm]
      \dot{y}_{\rm R} = x_{\rm R} + a y_{\rm R} \\[0.1cm]
      \displaystyle
      \dot{z}_{\rm R} = b -c z_{\rm R} + \frac{x_{\rm R} z_{\rm R}}{\rho}
    \end{array}
  \right.
\end{equation}
In that case, the actuating signal $u$ is designed with the term
\begin{equation}
  {\cal L}_{f_{\rm R}}^3 h = - b \rho + (a^2 - 1) x + a (a^2 -2) y  
        + \rho (c - a) z - xz  \, . 
\end{equation}
Using a coupling strength $\Lambda \in [-30, -5]$, we are able to control the 
R\"ossler dynamics to a chaotic solution of the H\'enon-Heiles system as 
plotted in Fig.\ \ref{henoros}(a). The Lissajous curve obtained by plotting 
$y_{\rm R}$ as a function of $x_{\rm H}$ is a sharp straight line 
($\overline{\epsilon}_{\rm L} < 10^{-4}$), showing the synchronization between 
the measured variables: this is a strong generalized synchronization. The 
different plane projections of the three-dimensional subspace related to the 
R\"ossler system reveals the chaotic nature of the controlled trajectory.

The Poincar\'e section of the dynamics for different sets of initial conditions
[Fig.\ \ref{henoros}(b)] reveals a chaotic sea with various quasi-periodic
islands. It also reveals that the invariant set presents some self-crossings
--- characterized by some overlap in the Poincar\'e section --- meaning that
the dynamics cannot be embedded within a three-dimensional space. The sections
are apparently non-conjugated. Nevertheless, each quasi-periodic island is
recovered in this Poincar\'e section. Those which are out of the 
$V_{y_{\rm HH}} = 0$ axis in the original section [Fig.\ \ref{henhei}(b)], 
their number is divided by two. The dissipativeness of the R\"ossler system
breaks the time reversibility and the symmetry observed in the original 
Poincar\'e section [Fig.\ \ref{henhei}(b)] is modded out. The overlaps are
similar to those observed when an image of a system with symmetry is 
constructed \cite{Mir93,Let01}. The maps are, therefore, conjugated, modulo 
the rotation symmetry inherent to the H\'enon-Heiles system. We have, therefore, a type-{\sc ii} generalized synchronization.

\subsection{Saito driving R\"ossler}

\begin{figure}[ht]
  \centering
  \includegraphics[width=0.46\textwidth]{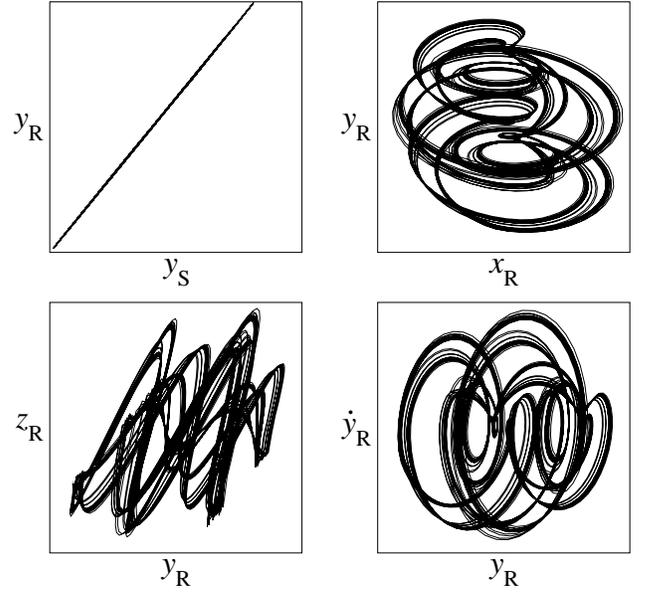} \\
  (a) Plane projections of the state portrait \\[0.2cm]
  \includegraphics[width=0.30\textwidth]{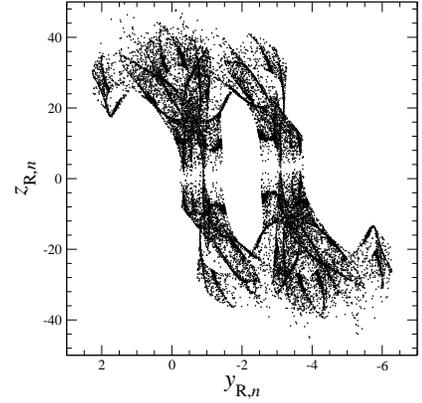} \\
  (b) Poincar\'e section \\[-0.2cm]
  \caption{Chaotic behavior produced by the response R\"ossler system 
driven by the four-dimensional Saito system (\ref{saiteq}). Parameter values
as in Figs.\ \ref{normros} and \ref{saitodyn}. Control parameter value:
	$\Lambda = - 20$ and $\overline{\epsilon}_{\rm L} < 10^{-3}$.}
  \label{rosaitoatt}
\end{figure}

Let us thirdly consider the case where the drive system is the 4D Saito system 
(\ref{saiteq}) and the response system is the 3D R\"ossler system
(\ref{Ros76}): the drive system has thus a greater dimension than the response 
system, and, by definition, only generalized synchronization can be obtained. To 
have the variable $y_{\rm S}$ of the Saito system within the same range of 
fluctuations as the variable $y_{\rm R}$ of the R\"ossler system, we applied 
the affine transformation $\tilde{y}_{\rm S} = 2 y_{\rm S} - 2$. 
We can get  synchronization between the two measured variables by 
applying the actuating signal $u$ with a coupling strength $\Lambda = -20$ as 
shown by the straight  Lissajous curve in Fig.\ \ref{rosaitoatt}(c). 

The Lissajous curve [Fig.\ \ref{rosaitoatt}(a], top left] between the measured
variables is a straight line ($\overline{\epsilon}_{\rm L} < 10^{-3}$): this is
a strong generalized synchronization. The plane projections of the response
dynamics resemble to the Saito dynamics [Fig.\ \ref{saitodyn}(a)]. When a 
Poincar\'e section is computed, we got a structure very similar to the original
one: they are at least $\epsilon$-conjugate. This is a type-{\sc ii} 
generalized synchronization. The response dynamics is necessarily 
four-dimensional and should be embedded in the space 
$(x_{\rm R}, y_{\rm R}, z_{\rm R}, \dot{y}_{\rm R})$, for instance. The 
Poincar\'e section is slightly different [compare Fig.\ \ref{rosaitoatt}(b)
with Fig.\ \ref{saitodyn}(b)]: the latter was rather homogeneously visited. At the same time, the section of the response attractor R\"ossler system presents some
folds that could be a signature of the dissipativeness of its dynamics. 
Unfortunately, the lack of robust topological characterization of dynamics
whose dimension is greater than 3 prevents a more conclusive answer. 

\subsection{R\"ossler driving Saito}

\begin{figure}[ht]
  \centering
  \includegraphics[width=0.46\textwidth]{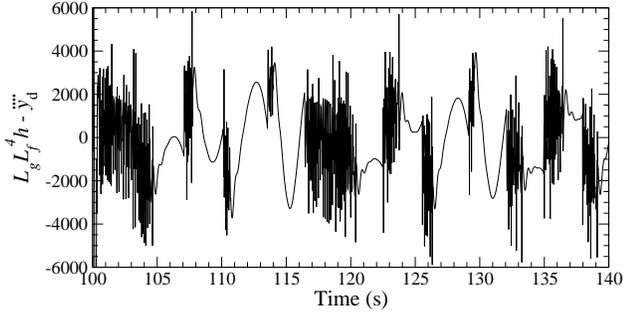} \\[-0.3cm]
  \caption{Time series of the term $({\cal L}_f^4h - \ddddot{y}_{\rm \! R})$.
Control parameter value: $\Lambda = - 20$ and 
$\overline{\epsilon}_{\rm L} < 0.002$.  }
  \label{saitorosLf4h}
\end{figure}

We now turn on the case where the 3D R\"ossler system drives the 4D Saito 
system. The dimension of the drive system is smaller than the dimension of the 
response system. The two measured variables are still $y_{\rm R}$ and 
$y_{\rm S}$. As investigated in a previous work \cite{Min23}, the actuating 
signal $u$ given by Eq.(\ref{actsig4D}) is applied to the derivative 
$\dot{w}_{\rm S}$. The term $({\cal L}_f^4h - \ddddot{\xi}_{\rm \! d})$ with 
$\xi_{\rm d}=y_{\rm R}$ is omitted due to large fluctuations observed (Fig.\ 
\ref{saitorosLf4h}) and, consequently, only the stabilizing terms are left in 
the actuating signal.

\begin{figure}[ht]
  \centering
  \includegraphics[width=0.46\textwidth]{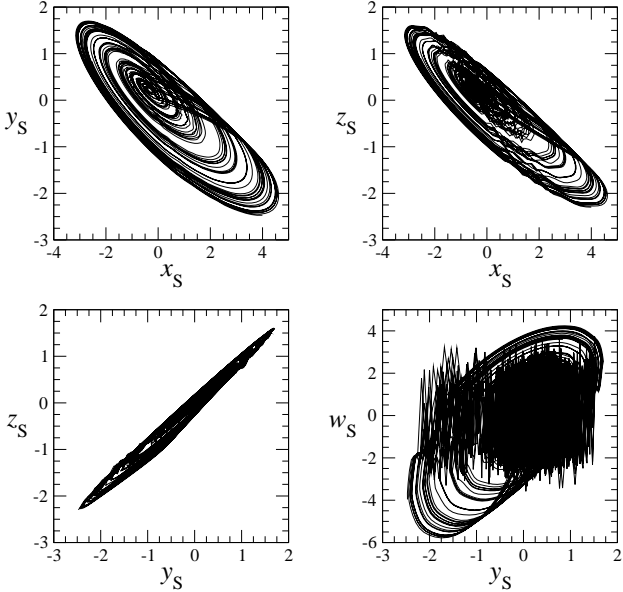} \\[-0.3cm]
  \caption{Response Saito dynamics to the drive R\"ossler system.
Same parameter values as in Fig.\ \ref{saitorosLf4h}.}
  \label{saitorosponse}
\end{figure}

When $\Lambda = - 20$, the Lissajous curve $y_{\rm R}$-$y_{\rm S}$ (not 
shown) is a sharp straight line: applying the actuating signal $u$ stabilizes 
the Saito dynamics into the drive R\"ossler one (Fig.\ \ref{saitorosponse})
and there is a generalized synchronization since the
three other variables, $x_{\rm S}$, $z_{\rm S}$, and $w_{\rm S}$, do not 
provide such a straight line but rather a complex figure resembling the 
R\"ossler attractor (not shown). Plane projections of the response dynamics
(Fig.\ \ref{saitorosponse}) exhibit a R\"ossler dynamics, which is quite smooth
in the $x_{\rm S}$-$y_{\rm S}$ plane and become more and more blurred with 
large fast fluctuations when the
variables used to span the projection are far from the $y_{\rm S}$ variable 
in the observability path for the Saito system [Fig.\ \ref{flusaito}(b)]. Thus,
the $y_{\rm S}$-$w_{\rm S}$ projection (Fig.\ \ref{saitorosponse}) shows large 
fluctuations blurring the structure of the R\"ossler attractor: such a feature 
directly results from the piecewise linear function $\Phi_w$ of the Saito system
(\ref{saiteq}). It can be asserted that the R\"ossler attractor projected 
in the subspace $\mathbb{R}^3 \left( \displaystyle x_{\rm S},
y_{\rm S}, z_{\rm S} \right)$ is topologically equivalent to the original 
R\"ossler attractor.

\begin{figure}[ht]
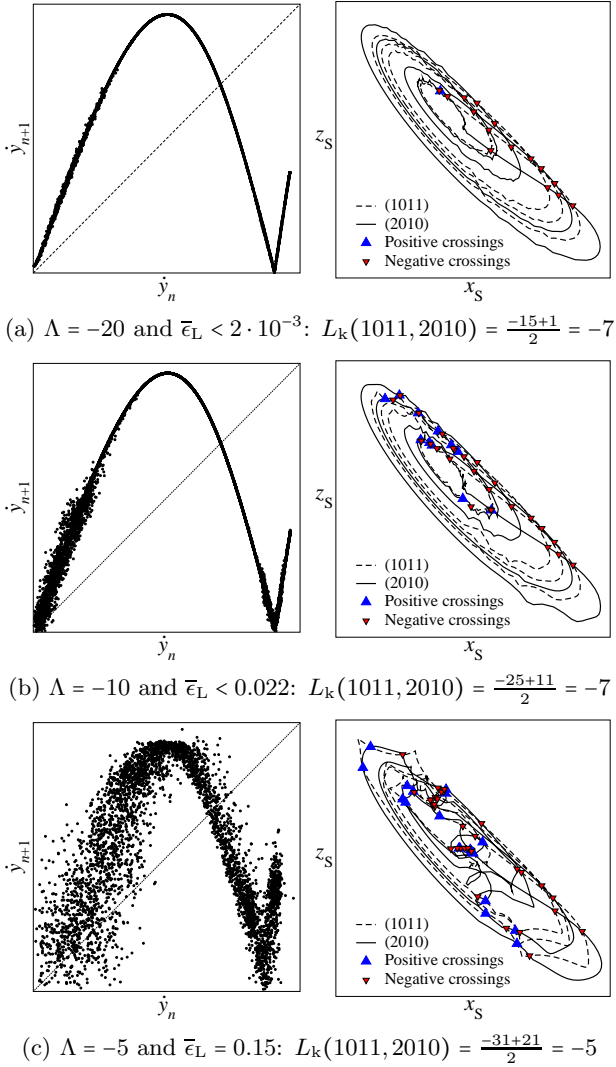

  \centering
  \begin{tabular}{cc}
    \includegraphics[width=0.22\textwidth]{saitorosmapL20.eps} &
    \includegraphics[width=0.22\textwidth]{saitorsupo_xzy_L20.eps} \\
	  \multicolumn{2}{c}{(a) $\Lambda  = - 20$ and 
	  $\overline{\epsilon}_{\rm L} < 2 \cdot 10^{-3}$: 
	  $L_{\rm k} (1011,2010) = \frac{-15 + 1}{2} = -7$} \\[0.2cm]
    \includegraphics[width=0.22\textwidth]{saitorosmapL10.eps} &
    \includegraphics[width=0.22\textwidth]{saitorsupo_xzy_L10.eps} \\
	  \multicolumn{2}{c}{(b) $\Lambda  = - 10$ and 
          $\overline{\epsilon}_{\rm L} < 0.022$:
	  $L_{\rm k} (1011,2010) = \frac{-25 + 11}{2} = -7$} \\[0.2cm]
    \includegraphics[width=0.22\textwidth]{saitorosmapL05.eps} &
    \includegraphics[width=0.22\textwidth]{saitorsupo_xzy_L05.eps} \\
	  \multicolumn{2}{c}{(c) $\Lambda  = - 5$ and 
          $\overline{\epsilon}_{\rm L} = 0.15$:
	  $L_{\rm k} (1011,2010) = \frac{-31 + 21}{2} = -5$} \\[-0.2cm]
  \end{tabular}
  \caption{First-return maps (left column) produced by the Saito system in
response to the R\"ossler system, and linking between the two period-4 
orbits (1011) and (2010) (right column) for three values of $\Lambda$.
Same parameter values as in Fig.\ \ref{saitorosLf4h}.}
  \label{saitrosmupo}
\end{figure}

We constructed the first-return map in the $y_{\rm S}$-induced differential
embedding by using the Poincar\'e section
\begin{equation}
  {\cal P}_{\rm S} = 
  \left\{
    \displaystyle 
	\left( \displaystyle \dot{y}_n, \ddot{y}_n \right) \in \mathbb{R}^2
	~|~ y_n = 0.21, \dot{y}_n < 0
  \right\} \, . 
\end{equation}
The corresponding first-return map is a bimodal map with a slightly thick
left increasing branch [Fig.\ \ref{saitrosmupo}(a), left]. With a good 
accuracy, this map can be considered as topologically conjugated to the 
bimodal map produced by the drive R\"ossler system [Fig.\ \ref{normros}(b)].
We extracted the population of periodic orbits from this map and computed the
linking number between the two of them.
We used the three-dimensional space $\mathbb{R}^3 (x_{\rm S},z_{\rm S},
y_{\rm S})$ discarding the $w_{\rm S}$-variable due to its jittery aspect 
induced by the switch function $\Phi (w_{\rm S})$ and we inverted the 
$y_{\rm S}$- and $z_{\rm S}$-axes for getting negative linking numbers as in 
the R\"ossler attractor \cite{Let95a}. Indeed, replacing a right-handed 
trihedron with a left-handed one inverts the sign of oriented crossings 
\cite{Let96d}. We thus obtained the linking number $L_{\rm k} (1011,2010) = -7$ 
as found in the R\"ossler attractor (not shown); the response dynamics can, therefore, be embedded within a three-dimensional space as the drive R\"ossler 
dynamics. We thus have a strong generalized synchronization with topologically 
conjugate first-return maps and a response dynamics with the same 
dimensionality as the drive dynamics. This is a type-{\sc i} generalized 
synchronization is observed for $\Lambda \in [-20,-10]$. 

\begin{figure}[ht]
  \centering
  \includegraphics[width=0.45\textwidth]{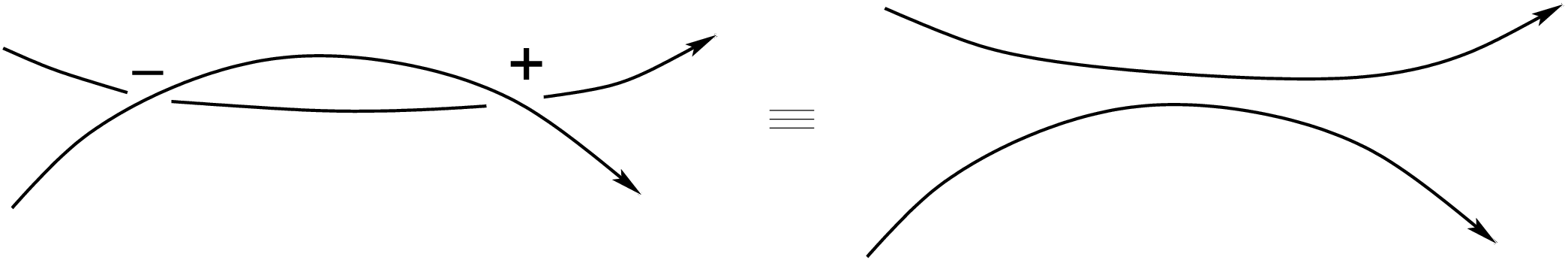} \\[-0.2cm]
  \caption{Removal of a pair of one negative and one positive crossings 
according to the type {\sc ii} Reidemesiter move.}
  \label{reidmove2}
\end{figure}

When $\Lambda = - 10$, the Lissajous curve is still a straight 
line, but the dynamics already shows some additional oscillations around the
expected drive dynamics as revealed by the thickness of the left increasing 
branch and the right end of the decreasing branch [Fig.\ \ref{saitrosmupo}(b)
left]. Instead of the 14 negative crossings counted in the R\"ossler attractor, 
there are now 25 negative and 11 positive crossings, which leads to a sum of 
$-14$, that is, $L_{\rm k} (1011,2010) = -7$, as in the R\"ossler dynamics. 
Each of the positive crossings can be paired with a negative one, and be 
removed according to the type {\sc ii} Reidemeister move (Fig.\ 
\ref{reidmove2}) \cite{Ale26b,Rei27}. In other words, the trajectory is now 
within an attractor with a given thickness --- one would say that the maps are 
$\epsilon$-conjugate, leading to a type-{\sc ii} generalized synchronization, 
that is, a weak generalized synchronization for which the embedding 
dimension of the response attractor is equal to the one of the drive invariant
set and for which the first-return maps are conjugated. Note that it
does not modify the linking number, that is, its description by a 
template \cite{Let18c}. We found that $\Lambda = - 10$ is the threshold value 
below which the response dynamics is topologically equivalent to the drive one 
since for $\Lambda = - 9$, there are two positive crossings between our 
period-4 orbits, which cannot be paired and, consequently, would induce a 
self-crossing  of the manifold on which the trajectory is structured (see 
below).

For $\Lambda = -5$, the control law is less efficient, taking more time to 
bring 
the controlled trajectory to the desired one. As a consequence, the response
dynamics is a combination of the drive dynamics and of the autonomous response
dynamics [Fig.\ \ref{saitrosmupo}(c), left]. The combination effect of these two dynamics is well-observed in the $x_{\rm S}$-$y_{\rm S}$ plane 
projection of the two orbits (1011) and (2010) whose jittery aspects clearly
appear [Fig.\ \ref{saitrosmupo}(c), right]. There are 21 positive and 31 
negative crossings, leading to a sum of $-10$: the linking number is therefore
$L_{\rm k} (1011,2020) = -5$. The topological properties are no longer 
preserved. In fact, contrary to what was observed for $\Lambda = - 10$, the 
positive crossings cannot be paired with one negative crossing by using a type {\sc ii} Reidemeister move (an isotopy, that is, a deformation without any 
cutting \cite{Tuf92}), which induces instead an actual
rotation of the trajectory around itself. To avoid the so-induced 
self-crossings of the manifold to itself, it is therefore necessary to 
increase the embedding dimension (with a false nearest neighbors 
algorithm \cite{Cao97a}, we found $d_{\rm E} = 4$ for $\Lambda = -5$ while it is
$d_{\rm E} = 3$ for $\Lambda = - 10$). The response dynamics is, therefore no
longer topologically equivalent to the drive dynamics. This is qualified as 
weak generalized synchronization: the maps are not topologically conjugate and
the dimension is not preserved. There is a W generalized synchronization
(see Table \ref{typeGS}).

\subsection{R\"ossler driving H\'enon-Heiles}https://www.overleaf.com/project/64aececfab7e7727129749e5

Here, the purpose is to drive a conservative system with a dissipative one. 
We cannot retain the Sprott A system since it is impossible to design
a global flat control law for it. We therefore used the four-dimensional 
H\'enon-Heiles system.

\begin{figure}[ht]
  \centering
  \includegraphics[width=0.46\textwidth]{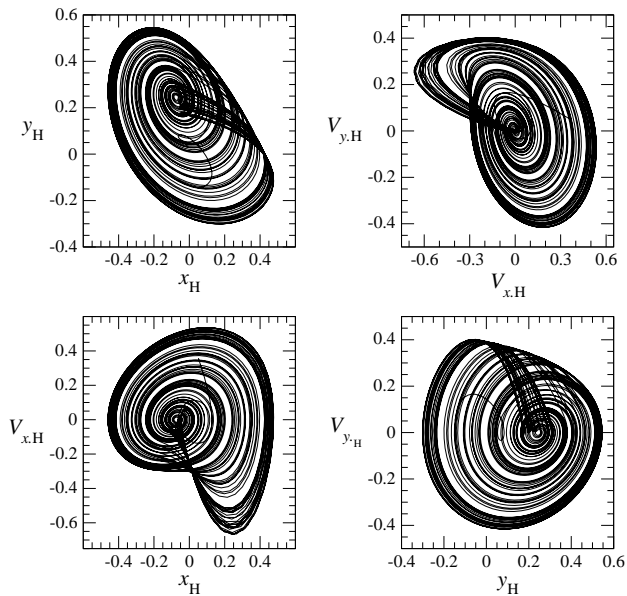} \\[-0.3cm]
  \caption{Response H\'enon-Heiles dynamics to the drive rescaled R\"ossler 
system. Parameter values: $a = 0.452$, $b = 2$, $c = 4$, $\rho = 0.1$, and 
$\Lambda = -1$ and $\overline{\epsilon}_{\rm L} = 10^{-3}$.}
  \label{rosheiles}
\end{figure}

To have $x_{\rm R}$ and $y_{\rm R}$ of the same order of magnitude as
$x_{\rm H}$ and $y_{\rm H}$, respectively, we used the rescaled R\"ossler
system (\ref{rescaros})
as the drive system
in which the measured variables are $x_{\rm R} + x_{\rm s}$ and 
$y_{\rm R} + y_{\rm s}$ where $x_{\rm s} = -0.1$ and $y_{\rm s} = 0.3$. 
The controlled H\'enon-Heiles system thus behaves as a dissipative R\"ossler
system (Fig.\ \ref{rosheiles}). The response four-dimensional conservative 
H\'enon-Heiles system now produces a three-dimensional dissipative dynamics.
For similar reasons to the previous cases, two pairs of variables, 
$(x_{\rm R}, x_{\rm H})$ and $(y_{\rm R}, y_{\rm H})$ are now synchronized;
nevertheless, the variable $z_{\rm R}$ is not simply related to any variable
of the H\'enon-Heiles system, and, consequently, this is still a generalized
synchronization. For $- 0.04 \geqslant \Lambda \geqslant -49$, the generalized
synchronization is strong with a trajectory embedded in a three-dimensional 
space and a first-return map which is topologically conjugate to the R\"ossler 
one: this is a type-{\sc i} generalized synchronization. For smaller 
$\Lambda$-values, the trajectory of the response system is 
ejected to infinity. For larger $\Lambda$-values, the response trajectory 
starts to be slightly jittery, fluctuating around the expected trajectory and 
with a state portrait shifted towards larger $x_{\rm H}$-values and smaller 
$y_{\rm H}$-values. We do not have $d_{\rm r} > d_{\rm d}$ with conjugate maps
because the response system has a larger dimension than the drive system, 
that is, an additional dimension to fluctuate without too strong constraint
from the drive system.

\section{Conclusion}
\label{conc}

Generalized synchronization is a very rich type of synchronization, mainly when 
occurring between systems whose governing equations are very different. We 
demonstrated that using a flat control law designed by  feedback linearization, 
it is possible to systematically get  such a generalized synchronization 
independently from the nature (dissipative or conservative) and dimensionality 
of the coupled systems. The condition for getting such synchronization is the 
same as for having a flat system: the response system must have a variable 
providing a global observability of the state space and, by duality, there is 
one derivative providing a global controllability. In addition, it is required 
that the $d$-dimensional response system is such that only the $d$th Lie 
derivative contributes to the actuating signal. If these conditions are 
fulfilled, then the response system can exhibit generalized synchronization 
with any driving dynamics: it might be necessary that the measured drive 
variable and its derivatives stay within the same variation range as those of 
the response system, which can be achieved with a simple time renormalization 
technique. Another technique, termed dilatation \cite{Ros92,Pol21}, can be used 
too although is more sophisticated and it is currently under investigation. We 
thus showed that we are able to get generalized synchronization between a 
conservative four-dimensional H\'enon-Heiles and a dissipative R\"ossler 
systems, for instance.

We thus distinguished four types of generalized synchronization by using a mean
thickness for the Lissajous curve, the minimal embedding dimension of the 
response dynamics, the topological conjugacy between the first-return maps
or Poincar\'e sections, and the nature of the map between the drive and the
response asymptotic invariant sets. Typically, when the mean thickness remains
below the threshold value of 0.05, we qualified the generalized synchronization
as being strong: the minimal embedding dimension of the response dynamics is 
equal to the embedding dimension of the drive dynamics. When the first-return 
maps or Poincar\'e sections are conjugate (or $\epsilon$-conjugate), we stay 
that this is type-{\sc i} (type-{\sc ii}) generalized synchronization. For 
larger thickness, the generalized synchronization is weak, and the map between
the drive asymptotic invariant set and the response one may be multivalued
(type-{\sc iii}) or even discontinuous (type-{\sc iv}). Generalized 
synchronization is thus ranked from the strongest (type-{\sc i}) to the weakest 
(type-{\sc iv}).

Using two structurally different systems as the drive and the response systems,
a flat control law applied to the response system allows when the coupling 
strength is sufficiently large to get a type-{\sc i} generalized 
synchronization. Decreasing the coupling strength, the generalized 
synchronization becomes weaker and, generally, the four types here introduced
can be distinguished. Obviously, when $d_{\rm d} > 3$, we face to the problem
of getting an accurate characterization of the dynamics and tools to check the
conjugacy of the Poincar\'e sections remains to be developed for a better
quantification of the different types of generalized synchronization.

Note that when the flat controlled system supports sufficiently large coupling
value, we are ensured to get a type-{\sc i} generalized synchronization, that
is, the response dynamics is equivalent to the drive one. Such a result 
reflects the great efficiency of using a flat control law.
With the help of flat coupling, we systematized a technique for getting 
generalized synchronization. The success of the approach only depends on the
possibility of designing a flat control law for the response system. The 
actuating signal can carry any dynamics until its oscillations are
of the same magnitude as the response system's measured variable.
Varying the coupling strength, it is easy to observe the four types of 
generalized synchronization here distinguished, confirming, if yet necessary,
that such synchronization is rich indeed. Depending on the task considered, the
response dynamics can be tuned from the isolated response dynamics to the 
drive one, with an intricated combination of these two dynamics for 
intermediate values of the coupling strength. Such a technique could be 
particularly interesting to impose a given dynamics to a network without 
changing the nature (the structure of the governing equations) of any node.

\acknowledgments
IL, ISN, and CL acknowledge support from the Spanish Ministerio de Ciencia e 
Innovaci\'on under Project PID2020-113737GB-I00. CL wishes to thank Witold 
Respondek for fruitful discussions. We thank Arkady Pikovsky for 
suggesting applying flat control to a two-dimensional oscillator with a chaotic
system. CL wishes to thank Ulrich Parlitz for constructive exchanges. LM 
gratefully acknowledges the support of the ``Hundred Talents'' program of the 
University of Electronic Science and Technology of China.

\bibliography{SysDyn}

\end{document}